\documentclass[11pt]{article} 
\usepackage{fullpage}
\usepackage[mathscr]{eucal}
\usepackage{xspace}
\usepackage{amsmath}
\usepackage{amssymb}
\usepackage{latexsym}

\begin{document}

\newcounter{casecounter}
\newenvironment{mycases}%
{\begin{list}
    {\bfseries Case \arabic{casecounter}:}
    {
      \setlength{\itemindent}{0in}
     \usecounter{casecounter}}}
 {\end{list}}

\newtheorem%
     {theorem}{Theorem}[section]
\newtheorem%
     {corollary}[theorem]{Corollary}
\newtheorem%
     {proposition}[theorem]{Proposition} 
\newtheorem%
     {lemma}[theorem]{Lemma} 

\newtheorem%
     {exampleAux}[theorem]{Example}
\newenvironment%
     {example}{\begin{exampleAux}\upshape}{\qed\end{exampleAux}}

\newtheorem%
     {examplesAux}[theorem]{Examples} 
\newenvironment%
     {examples}{\begin{examplesAux}\upshape}{\qed\end{examplesAux}}

\newtheorem{definition}[theorem]{Definition}

\newtheorem%
     {constructionAux}[theorem]{Construction} 
\newenvironment%
     {construction}{\begin{constructionAux}\rm}{\end{constructionAux}}

\newcommand{\CASE}[1]{{\it Case\/} #1.}

\def\proofstring{{\sc Proof.\ \ }}

\newenvironment%
     {proof}{\noindent\proofstring }{\qed}

\def\qed{\hfill{\boxit{}}
  \ifdim\lastskip<\medskipamount \removelastskip\penalty55\medskip\fi}
\long\def\boxit#1{\vbox{\hrule\hbox{\vrule\kern3pt
                  \vbox{\kern3pt#1\kern3pt}\kern3pt\vrule}\hrule}}

\newenvironment%
     {genass}%
        {\medbreak\noindent{\bf General Assumption.\enspace}\it}%
        {\ifdim\lastskip<\medskipamount \removelastskip\penalty55\medskip\fi}


\newcommand {\boxfigure}[1]%
   {\framebox[\textwidth]{%
    \parbox {0.99\textwidth}
                {{#1}\vspace {0cm}\hfill}}}

\newcommand {\boxfigureone}[1]%
   {\framebox[\textwidth]{%
    \parbox {0.90\textwidth}
                {{#1}\vspace {0cm}\hfill}}}

\newcommand{\gefig}[3]
        {\begin{figure}[htb] 
        \begin{center}%
        \mbox{}
        {\psfig{figure=#1,width=0.60\textwidth}}
        \mbox{}
        \end{center}
        \caption{{#2}\label{#3}} 
        \end{figure}}

\newcommand{\hangif}[1]{\raisebox{-1ex}{\hspace{2em}
                                        \makebox[1.0em][l]{if}
                                        \parbox[t]{320pt}{#1}}}


\def\A{{\cal A}} \def\B{{\cal B}} \def\C{{\cal C}} \def\D{{\cal D}}
\def\E{{\cal E}} \def\F{{\cal F}} \def\G{{\cal G}} 
\def\I{{\cal I}} \def\J{{\cal J}} \def\K{{\cal K}} \def\L{{\cal L}}
\def\M{{\cal M}} \def\N{{\cal N}} \def\O{{\cal O}} \def\P{{\cal P}}
\def\Q{{\cal Q}} \def\R{{\cal R}} \def\S{{\cal S}} \def\T{{\cal T}}
\def\U{{\cal U}} \def\V{{\cal V}} \def\W{{\cal W}} \def\X{{\cal X}}
\def\Y{{\cal Y}} \def\Z{{\cal Z}}

\newcommand{\dd}[2]{#1_1,\ldots,#1_{#2}}      

\newcommand{\set}[1]{\{\,#1\,\}}
\newcommand{\eset}{\emptyset}
\newcommand\bigset[1]{ \Bigl\{ #1 \Bigr\} }   
\newcommand\bigmid{\ \Big|\ }

\newcommand{\incl}{\subseteq}           
\newcommand{\incls}{\supseteq}          

\newcommand{\col}{\colon}

\newcommand{\NP}{{\rm NP}}              
\newcommand{\GI}{{\rm GI}}              
\newcommand{\PIPEETWO}{\Pi^{{\rm P}}_2} 
\newcommand{\SIGPEETWO}{\Sigma^{{\rm P}}_2}     
\newcommand{\PSPACE}{{\rm PSPACE}}      
\newcommand{\PTIME}{{\rm P}}            
\newcommand{\EXPTIME}{{\rm EXPTIME}}    

\newcommand{\angles}[1]{\langle#1\rangle}       

\newcommand{\impl}{\Rightarrow}   


\newcommand{\OnlyIf}{\lq\lq$\Rightarrow$\rq\rq\ \ }   
\newcommand{\If}{\lq\lq$\Leftarrow$\rq\rq\ \ }        

\newcommand{\Base}{{\bf Base Case:\ }}
\newcommand{\Induction}{{\bf Induction Step:\ }}

\newcommand{\quotes}[1]{\lq\lq#1\rq\rq}         
\newcommand{\wrt}{with respect to\xspace}              
\newcommand{\WLOG}{w.l.o.g.}                    
\newcommand{\ie}{i.e.}                          
\newcommand{\eg}{e.g.}                          
\renewcommand{\hom}{homo\-mor\-phism}

\newcommand{\eat}[1]{}

\newcommand{\comment}[1]{\noindent{\it COMMENT:} {\it #1}}

\newcommand{\bagt}[1]{t(#1)} 
\newcommand{\bagc}[1]{c(#1)} 
\newcommand{\bagv}[1]{v(#1)} 

\newcommand{\equivNum}[1]{|[#1]|} 

\newcommand{\bigforall}[1]{\underset{#1}{\bigvee\hspace{-2.33ex}\rule[0.6ex]{1.2ex}{0.25ex}}} 

\newcommand{\myvpace}{\rule[-2ex]{0ex}{5ex}}

\newcommand{\eqNum}[1]{\stackrel{#1}{=}}

\newcommand{\alphaSum}[1]{(\sum){}^{#1}}
\newcommand{\alphaSumI}[2]{(\sum_{#2}){}^{#1}}
\newcommand{\add}[1]{+_{#1}}
\newcommand{\subtract}[1]{-_{#1}}

\newcommand{\etpl}{\epsilon}          
\newcommand{\nequiv}[1]{\equiv_{#1}}

\newcommand{\locequiv}{\equiv_{\mbox{\scriptsize\sl loc}}}

\newcommand{\bagExpand}[2]{#1_{\otimes #2}}
\newcommand{\qtimes}{\otimes}
\newcommand{\proj}[1]{\tilde{#1}}       

\newcommand{\compose}{\circ}
\newcommand{\Map}{{\it Map}}

\newcommand{\per}{{\bf .}}              
\newcommand\osp{\;}                     
\newcommand\osps[1]{{\osp{#1}\osp}}     

\newcommand{\id}{{\it id}}              

\newcommand{\bagEqCon}{\subseteq_{\tt bag}^{\mbox{\tiny =}}}
\newcommand{\bagEquiv}{\equiv_{\tt bag}} 

\newcommand{\bsSubOld}{\subseteq_{\tt bs}} 
\newcommand{\bsSub}{\subseteq_{\tt bs}^{\mbox{\tiny =}}}    

\newcommand{\leqCon}{\bsSubOld}
\newcommand{\eqCon}{\bsSub}

\newcommand{\bagNum}[2]{|#2|_{#1}}     

\newcommand{\bsEquiv}{\equiv_{\tt bs}} 


\newcommand{\tpl}[1]{\bar{#1}}          
\newcommand{\ar}[2]{#1^{(#2)}}          
\newcommand{\SIG}{\Sigma}               

\newcommand{\type}{\tau}                

\newcommand\AND{\wedge}              
\newcommand\OR{\vee}                    
\newcommand\BIGOR{\bigvee}              
\newcommand{\qif}{\leftarrow}           

\newcommand{\having}{{\tt HAVING}}      
\newcommand{\union}{{\tt UNION}}        


\newcommand{\carr}[1]{\mbox{\sl carr}(#1)} 

\newcommand{\DB}[2]{{#1}^{#2}}          
\newcommand{\DBD}[1]{\DB{#1}{\D}}       
\newcommand{\DBE}[1]{\DB{#1}{\E}}       

\newcommand{\DBN}{\D_{\tpl N}}          

\newcommand{\bag}[1]{\{\!\!\{\,#1\,\}\!\!\}}    

\newcommand{\DBBAG}[2]{\bag{#1}^{#2}}   
\newcommand{\DBDBAG}[1]{\DBBAG{#1}{\D}} 

\newcommand{\MULT}[1]{\M(#1)}           

\newcommand{\bigbag}[1]{\Bigl\{\!\!\Bigl\{#1\Bigr\}\!\!\Bigr\}}    

\newcommand{\MIN}{\mbox{\sl min}}            
\newcommand{\MAX}{\mbox{\sl max}}            
\newcommand{\COUNT}{\mbox{\sl count}}        
\newcommand{\CNTD}{\mbox{\sl cntd}}          
\newcommand{\SUM}{\mbox{\sl sum}}            
\newcommand{\PROD}{\mbox{\sl prod}}          
\newcommand{\AVG}{\mbox{\sl avg}}            
\newcommand{\PRTY}{\mbox{\sl parity}}        
\newcommand{\MDN}{\mbox{\sl median}}         
\newcommand{\TOPTWO}{\mbox{\sl top$2$}}      
\newcommand{\TOPK}{\mbox{\sl top$K$}}        
\newcommand{\BOTTWO}{\mbox{\sl bot$2$}}      
\newcommand{\STDV}{\mbox{\sl stdev}}          

\newcommand{\CLASS}[2]{[#1]_{#2}}       

\newcommand{\eqvq}{\sim_{q}}            
\newcommand{\eqv}{\sim}                 
\newcommand{\classq}[1]{\CLASS{#1}{q}}  
\newcommand{\class}[1]{[#1]}            

\newcommand{\deqv}{\approx}             
\newcommand{\dclass}[1]{\CLASS{#1}{\deqv}}      

\newcommand{\card}[1]{|#1|}             
\newcommand{\bigcard}[1]{\Bigl|#1\Bigr|}                
\newcommand{\CARD}[1]{\mbox{\sl card}#1}     

\newcommand{\Ker}[2]{#1_{#2}}           

\newcommand{\core}[1]{\breve{#1}}               
\newcommand{\extinst}[1]{\hat{#1}}      

\newcommand{\ASS}[2]{\Gamma(#1,#2)}
\newcommand{\ASSqD}{\ASS q\D}
\newcommand{\ASStpl}[3]{\Gamma_{#3}(#1,#2)}
\newcommand{\ASSd}[2]{\ASStpl{#1}{#2}{\tpl d}}
\newcommand{\ASSqDd}{\ASStpl{q}{\D}{\tpl d}}


\newcommand{\dop}{\succeq}              
\newcommand{\spod}{\prec}               
\newcommand{\sdop}{\succ}               

\newcommand{\phiuv}{\phi_{\tpl u}^{\tpl v}}

\newcommand{\rat}{{\bf Q}}              
\newcommand{\ratnozero}{{\bf Q^\pm}}    
\newcommand{\nat}{{\bf N}}              
\newcommand{\intg}{{\bf Z}}             
\newcommand{\real}{{\bf R}}             


\newcommand{\qcount}{\Gamma}            

\newcommand{\HOM}{\mbox{\sl Hom}}

\newcommand{\ptrn}{\P}

\newcommand{\comp}[1]{\tilde #1}        

\newcommand{\essnovar}[1]{{\sharp}_{\it v}(#1)}
\newcommand{\essnorel}[1]{{\sharp}_{\it r}(#1)}

\newcommand{\red}[1]{#1^{{\it r}}}      
\newcommand{\eq}[1]{#1^{=}}             

\newcommand{\intmodels}{\models_\intg}  
\newcommand{\ratmodels}{\models_\rat}   
\newcommand{\dommodels}[1]{\models_{#1}}   

\newcommand{\udist}[3]{{\it dist}^{\uparrow}_{#1}(#2,#3)}
\newcommand{\ddist}[3]{{\it dist}^{\downarrow}_{#1}(#2,#3)}
\newcommand{\udistc}[2]{\udist{C}{#1}{#2}}
\newcommand{\ddistc}[2]{\ddist{C}{#1}{#2}}

\newcommand{\GLL}{{\it gll}}
\newcommand{\GLLC}{{\it gll}_C}
\newcommand{\gll}[2]{\GLL_{#1}({#2})}
\newcommand{\LUL}{{\it lul}}
\newcommand{\LULC}{{\it lul}_C}
\newcommand{\lul}[2]{\LUL_{#1}(#2)}
\newcommand{\gllc}[1]{\gll{C}{#1}}
\newcommand{\lulc}[1]{\lul{C}{#1}}
\newcommand{\glll}[1]{\gll{L}{#1}}
\newcommand{\lull}[1]{\lul{L}{#1}}

\newcommand{\gamdown}{\gamma^{\downarrow}}
\newcommand{\gamup}{\gamma^{\uparrow}}

\newcommand{\YPLUS}{Y^{+}}
\newcommand{\YMIN}{Y^{-}}
\newcommand{\dmin}{d^{-}}
\newcommand{\dplus}{d^{+}}

\newcommand{\VIRT}[1]{{\it vc}_{#1}}    
\newcommand{\VIRTC}{\VIRT C}            
\newcommand{\virt}[2]{{\it vc}_{#1}(#2)}
\newcommand{\virtc}[1]{\virt{C}{#1}}    


\newcommand{\lin}[2]{\L_{#1}(#2)}       
\newcommand{\lind}[1]{\lin{D}{#1}}      

\newcommand{\varl}[2]{\L^{\it v}_{#1}(#2)}      
\newcommand{\varld}[1]{\varl{D}{#1}}    

\newcommand{\consl}[2]{\C\L_{#1}(#2)}   
\newcommand{\consld}[1]{\consl{D}{#1}}  

\newcommand{\ndcqd}{N_D({\core q},d)}   
\newcommand{\ndcqpd}{N_D({\core q'},d)} 

\newcommand{\q}[1]{q_{#1}}              
\newcommand{\ql}{\q{L}}                 
\newcommand{\qll}{(\ql)_L}              
\newcommand{\qp}[1]{q'_{#1}}            
\newcommand{\qpm}{\qp{M}}               
\newcommand{\qpmm}{(\qpm)_{M}}          

\newcommand{\cq}[1]{\core{q}_{#1}}      
\newcommand{\cql}{\cq{L}}               
\newcommand{\cqil}{\cq{iL}}             
\newcommand{\cqll}{(\cql)_L}            
\newcommand{\cqp}[1]{\core{q}'_{#1}}    
\newcommand{\cqpm}{\cqp{M}}             
\newcommand{\cqpjm}{\cqp{jM}}           
\newcommand{\cqpmm}{(\cqpm)_{M}}        

\newcommand{\wiso}{\sim}                

\newcommand{\phil}{\phi_L}              
\newcommand{\phim}{\phi_M}              

\newcommand{\sig}[1]{\sigma(#1)}        

\newcommand{\QV}{\Q_\V}                 
\newcommand{\QC}{\Q_\C}                 

\newcommand{\rel}[1]{{#1}^{\it rel}}    

\newcommand{\lnkd}{\leftrightarrow}     


\newcommand{\lexp}[1]{{#1}^{{\it lin}}} 

\newcommand{\expqll}{\BIGOR_L \ql}      
\newcommand{\expqpmm}{\BIGOR_M \qm}     


\newcommand{\inv}[1]{#1^{-1}}           

\newcommand{\restr}[2]{#1_{|#2}}        

\newcommand{\var}[1]{{\it var}(#1)}     

\newcommand{\NDVAR}[1]{{\it ndv}(#1)}   

\newcommand{\tsize}[1]{\tau(#1)}


\newcommand{\EXTDB}[2]{#1_#2}           

\newcommand{\equivv}{\equiv_\V}         
\newcommand{\equivvp}{\equiv_{\V'}}     

\renewcommand{\exp}[1]{\tilde #1}       

\newcommand{\unf}[1]{#1^{{\it u}}}      

\newcommand{\Count}{{\it c}}                    
\newcommand{\Sum}{{\it s}}                      
\newcommand{\Max}{{\it m}}                      
\newcommand{\Reg}{{\it r}}                      

\newcommand{\VC}{\V^\Count}                     
\newcommand{\VS}{\V^\Sum}                       
\newcommand{\VM}{\V^\Max}                       
\newcommand{\VR}{\V^\Reg}                       

\newcommand{\vc}{v^\Count}                      
\newcommand{\vs}{v^\Sum}                        
\newcommand{\vm}{v^\Max}                        
\newcommand{\vr}{v^\Reg}                        


\newcommand{\wclass}[1]{[#1]_{\it w}}           


\newcommand{\DW}{data warehouse}

\newcommand{\Corr}[1]{#1}
\newcommand{\RM}[1]{}

\newcommand{\sfq}{{\sf q}}
\newcommand{\ta}{{\sf ta}}
\newcommand{\salaries}{{\sf salaries}}

\newcommand{\Name}{{\it Name}}
\newcommand{\CourseName}{{\it Course\_Name}}
\newcommand{\JobType}{{\it Job\_Type}}
\newcommand{\Amount}{{\it Amount}}
\newcommand{\Sponsorship}{{\it Sponsorship}}
\newcommand{\ISF}{\mbox{\tt 'Govt.'}}


\newcommand{\BASE}{\mbox{\sf BASE}}
\newcommand{\Bag}{{\it Val}}
\newcommand{\BMap}{{\it VM}}
\newcommand{\arity}{\mbox{\sl ary}}
\newcommand{\scriptarity}{\mbox{\scriptsize\sl ary}}


\newcommand{\topstwo}{{\bf T}_2}        


\newcommand{\pos}[1]{#1_{+}}            

\newcommand{\todo}[1]{{\bf To DO: #1}}


\title{{Equivalences Among Aggregate Queries\\ with Negation}
\author{Sara Cohen%
        \thanks{The Hebrew University, Institute of 
                Computer Science, Jerusalem 91904, Israel. 
                Supported by Grant 96/01-1 from the 
                Israel Science Foundation.} 
\and    Werner Nutt%
        \thanks{School of Mathematical and Computer Sciences,
                Heriot-Watt University, Edinburgh EH14 4AS,
                Scotland, UK.
                Supported by Grant GR/R74932/01 from the 
                Engineering and Physical Sciences Research Council.}
\and    Yehoshua Sagiv${}^\ast$
}
}

\eat{
\numberofauthors{3}
\author{%
\alignauthor Sara Cohen \\
\affaddr{Institute of Computer Science} \\
\affaddr{The Hebrew University} \\
\affaddr{Jerusalem 91904, Israel} \\
\email{sarina@cs.huji.ac.il}
\alignauthor Werner Nutt \\
\affaddr{Department of Computing}\\
\affaddr{and Electrical Engineering} \\
\affaddr{Heriot-Watt University, Edinburgh EH14 4AS} \\
\affaddr{Scotland, UK} \\
\email{nutt@cee.hw.ac.uk} 
\alignauthor Yehoshua Sagiv \\
\affaddr{Institute of Computer Science} \\
\affaddr{The Hebrew University} \\
\affaddr{Jerusalem 91904, Israel} \\
\email{sagiv@cs.huji.ac.il}}
}
\date{}
\maketitle

\vspace{1ex}

\begin{abstract}
Query equivalence is investigated for disjunctive aggregate
queries with negated subgoals, constants and  comparisons.
A full characterization of equivalence is given for the aggregation
functions $\COUNT$, $\MAX$, $\SUM$, $\PROD$, $\TOPTWO$ and $\PRTY$.
A related problem is that of determining, for a given natural number
$N$, whether two given queries are equivalent over all databases with
at most $N$ constants. We call this problem {\em bounded
  equivalence.\/} A complete characterization of decidability of
 bounded equivalence is given.
In particular, it is shown that this problem is decidable 
for all the above aggregation
functions as well as for $\CNTD$ (count distinct) and $\AVG$.
For quasilinear queries (i.e., queries where predicates that occur
positively are not repeated) it is shown that equivalence can be
decided in polynomial time for the aggregation functions $\COUNT$,
$\MAX$, $\SUM$, $\PRTY$, $\PROD$, $\TOPTWO$ and $\AVG$.
A similar result holds for $\CNTD$ provided that a few
additional conditions hold.
The results are couched in terms of abstract characteristics of
aggregation functions, and new proof techniques are used.
Finally, the results above also imply that equivalence, under
bag-set semantics, is decidable for non-aggregate queries with
negation.
\end{abstract}




\section{Introduction}

The emergence of data warehouses and of decision-support systems has
highlighted the importance of efficiently processing aggregate
queries. 
In such systems the amount of data is generally large and aggregate
queries are used as a standard means of reducing the volume of the
data.
Aggregate queries tend to be expensive as they ``touch'' many items
while returning few.
Thus, optimization techniques for aggregate queries are a necessity.
Many optimization techniques, such as query rewriting, are based on
checking query equivalence.
For this purpose, a coherent understanding of the equivalence problem
of aggregate queries is necessary.

One of our main results in this paper is that equivalence 
is decidable for disjunctive
queries with comparisons and negated subgoals if they contain one of
the aggregation functions:
$\MAX$, $\TOPTWO$, 
$\COUNT$, $\SUM$, $\PROD$, or $\PRTY$.

A query that does not have negated subgoals is {\em positive}.
Equivalence of positive non-aggregate queries has been
studied extensively~%
\cite{Chandra:Merlin-Conjunctive:Queries-STOC,%
        Aho:Et:Al-Efficient:Optimization-TODS,%
        Sagiv:Yannakakis-Union:And:Difference-JACM,%
        Johnson:Klug-Optimizing:Conjunctive:Queries-SIAMJ,%
        Van:Der:Meyden-Linearly:Ordered:Domains-PODS,%
        Levy:Sagiv-Sem:Query:Opt-PODS}.
Furthermore, in~\cite{Levy:Sagiv-Queries:Independent:Updates-VLDB}
it has been shown that equivalence is decidable
for non-aggregate disjunctive queries with negation.
Syntactic characterizations of equivalences among aggregate queries
with the functions $\MAX$, $\SUM$, and 
$\COUNT$ have been given in~%
\cite{Nutt:Et:Al-Equivalences:Among:Aggregate:Queries-PODS,%
        Cohen:Et:Al-Rewriting:Aggregate:Queries-PODS}.
These results have been extended in~%
\cite{Grumbach:Rafanelli:Tininini-Querying:Aggregate:Data-PODS}
to queries with the functions $\PROD$ and $\AVG$, 
for the special case of queries that contain neither constants nor
comparisons.
Thus, there are results on the equivalence problem for 
{\em non\/}-aggregate queries {\em with\/} negation as well as for
{\em aggregate\/} queries {\em without\/} negation.
Equivalence of {\em aggregate\/} queries {\em with\/}
negated subgoals was dealt with for the first time in~\cite{Negation}. This
paper is a substantially revised and extended version
of~\cite{Negation}.\footnote{One of the proofs in~\cite{Negation} was
  incorrect and has been corrected in this version.}

Our decidability proofs rely on abstract properties of aggregation
functions.  We consider functions that are defined by means of
operations on abelian monoids.
Our proofs work out if the monoids are either idempotent or are
groups.
Functions of the first kind are $\MAX$ and $\TOPTWO$, 
functions of the second kind are $\COUNT$, $\SUM$, and 
$\PRTY$.

For these functions we reduce equivalence with
respect to all possible databases to equivalence over databases that
have at most as many constants as there are constants and variables in
the queries, 
a property which we call {\em local equivalence.\/}
We do not study local equivalence immediately, but rather 
the more general problem of {\em bounded equivalence.\/}
It consists of determining, given a nonnegative integer $N$ and two
queries, whether the queries return identical results over all
databases with at most $N$ constants. We give a complete characterization
of decidability of bounded equivalence.
In particular, we show that bounded equivalence 
 is decidable for queries with the
functions $\COUNT$, $\CNTD$, $\MAX$, $\SUM$, $\PROD$, $\AVG$,  
$\TOPTWO$ and $\PRTY$.

Finally, we consider the special case of {\em quasilinear queries,\/}
that is, queries where predicates that occur positively are not
repeated.
For quasilinear queries 
equivalence boils down to isomorphism, 
which can be decided in polynomial time.


\section{Aggregation Functions}
        \label{sec-agg:functions}

An aggregate query is executed in two steps.
First, data is collected from a database as specified by the
non-aggregate part of the query.  
Then the results are grouped into multisets (or bags), 
an aggregation function is applied to the multisets, and the aggregates
are returned as answers.

The queries that we consider in this paper contain the
aggregation functions
$\COUNT$ and $\CNTD$, which for a bag 
return the number of elements or distinct elements, respectively;
$\PRTY$, which returns~$0$ or $1$, depending on whether the number of
elements in the bag is even or odd; 
$\SUM$, $\PROD$ and $\AVG$, which return the sum, product, or average
of the elements of a bag;
$\MAX$, which returns the maximum among the elements of a bag;
and $\TOPTWO$, which returns a pair consisting of the two greatest
different elements of a bag.

The reader will notice in the course of the paper
that our results for $\MAX$ and $\TOPTWO$
immediately carry over to $\MIN$ and $\BOTTWO$,
which select the minimum or the two least elements out of a multiset
of numbers.
Moreover, our results for $\TOPTWO$ can easily be generalized to the
function $\TOPK$, which selects the $K$ greatest different elements.

Our arguments to prove decidability of
equivalence for certain classes of aggregate queries rely on the fact
that the aggregation functions take values in special kinds of abelian
monoids and are defined in terms of the operations of those monoids.
To make this formal, we will introduce the class of 
{\em monoidal aggregation functions\/} and two of its subclasses.
We will show that all of the above functions except $\CNTD$, $\PROD$ and 
$\AVG$ belong to one of these two subclasses.
In general, an {\em aggregation function\/} maps multisets 
of tuples of numbers to values in some structure, 
which in most cases consists again of numbers.
Here, we assume that the results of the aggregation are
elements of some abelian monoid.

An {\em abelian monoid\/} is a structure $(M,+,0)$ 
consisting of a set $M$ with an associative and
commutative binary operation, which we denote as \quotes{$+$},
and a neutral element, which we denote as 0.
If no confusion can arise, we identify a monoid with the set on
which it is defined and refer to $(M,+,0)$ simply as the monoid $M$.

An abelian monoid $M$ is {\em idempotent\/} if $a + a = a$ holds
for all $a \in M$, and $M$ is a {\em group\/} if for every
$a\in M$ there is a $b\in M$ such that $a + b = 0$.
The element $b$ is called the {\em inverse\/} of $a$ and is usually
denoted as $-a$.  
Instead of $a + (-b)$ we will usually write $a - b$.

\begin{examples}
Standard  abelian monoids are the set of integers 
$\intg$ and the set of rational numbers $\rat$,
with the binary operation of addition and the neutral element $0$. We
use $\ratnozero$ to denote the set of rational numbers without the element
0. Note that $\ratnozero$, with the binary operation
of multiplication and the neutral element $1$ is also an abelian monoid.
A~further example is the two-element group $\intg_2 = \set{0,1}$, 
where the addition satisfies $1+1 = 0$.  

By $\rat_\bot$  we denote the rational numbers augmented by a new
element $\bot$, which is less than any element in $\rat$.  
Then $\rat_\bot$ is an abelian monoid if the operation is selecting
the maximum of two numbers.
The neutral element is $\bot$.

A less common example is the monoid $\topstwo$, which is defined on
the set of pairs
\begin{eqnarray*}
        \topstwo  &:=&  \bigset{(d,e) \in \rat_\bot\times\rat_\bot \bigmid
                                d > e}
                      \,\cup\, 
                      \bigset{ (\bot,\bot) }.
\end{eqnarray*}
We denote the binary operation on $\topstwo$ as \quotes{$\oplus$}.
We define $(d_1,e_1) \oplus (d_2,e_2)$ as the pair $(d,e)$ that
consists of the two greatest different elements among 
$\set{d_1,e_1,d_2,e_2}$, provided this set has at least two elements,
and as the pair $(d,\bot)$ if the set consists only of the element~$d$.
For instance, we have that
       $(5,\bot)\oplus(2,1) = (5,2)$, that
       $(5,2)\oplus(5,1) = (5,2)$, and that
       $(5,\bot)\oplus(5,\bot) = (5,\bot)$.
Clearly, $(\bot,\bot)$ is the neutral element.
\end{examples}

If $(M,+,0)$ is an abelian monoid, we can extend the binary operation
to subsets of $M$ and to multisets over $M$ in a canonical
way---because of the associativity and commutativity of \quotes{$+$},
the order in which we apply the operation does not matter.
If $S$ is such a set or multiset, 
we denote the result of applying \quotes{$+$} to $S$
as $\sum_{a\in S} a$.

Many common aggregation functions are computed by first mapping the
elements of a multiset of tuples into an abelian monoid and then
combining the values obtained through the mapping by the monoid operation.

Later on in the paper, we will assume that aggregate queries range over
databases with constants from some set $\I$.
We call such a set a {\em domain\/} and assume that a linear ordering 
\quotes{$<$} is defined on its elements.  
For example, the integers $\intg$ and the rational numbers $\rat$ 
are such domains.
For our discussion of aggregation functions the ordering on the domains
is of no importance.

If $\I$ is a domain and $k$ is a nonnegative integer, 
we use $\I^k$ to denote the $k$-fold cartesian product of $\I$.
Thus, $\I^k$ consists of all $k$-tuples where the components are
elements of $\I$.
For the special case in which  $k=0$, 
the set $\I^0$ consists of a single element, 
called the empty tuple. 
When $k=1$, we will often omit the superscript. 
Hence, we use $\I$ to denote $\I^1$. 

Technically, we assume that there is a domain $\I$, 
a nonnegative integer $k$, 
a monoid $(M,+,0)$ and 
a function $f\col \I^k\to M$.  
Then the  {\em aggregation function\/} over $\I^k$ based on $f$ and
\quotes{$+$}, 
which maps multisets $B$ over $\I^k$ to elements of $M$, is denoted
as $\alpha_f^{+}$ and defined by 
\begin{eqnarray*}
        \alpha_f^{+}(B)  :=  \sum_{a\in B} f(a),
\end{eqnarray*}
for all bags $B$ over $\I^k$.
We say that $\alpha$ is a {\em monoid aggregation function\/}
if $\alpha = \alpha_f^{+}$ for some abelian monoid operation 
\quotes{$+$}.
In particular, we say that $\alpha_f^{+}$ is {\em idempotent\/}
or a {\em group aggregation function\/} 
if the underlying monoid is idempotent, or a group,
respectively.

\begin{examples}
Obviously, $\SUM$ and $\MAX$ 
are the unary aggregation functions over $\intg$ or $\rat$,
based on the identity mapping and on addition or 
the binary operation \quotes{$\max$,}
respectively. 
The functions $\COUNT$ and $\PRTY$ over $\intg^0$ arise
from the additive  
groups $\intg$ and $\intg_2$, respectively, by choosing as $f$ the mapping 
that maps every element  to the constant
$1$. Note that $\COUNT$ and $\PRTY$ are {\em nullary\/} aggregation
functions. Therefore, the domain over which they are defined consists
only of the empty tuple.
We obtain the aggregate $\TOPTWO$ over $\rat$ by choosing
the monoid $\topstwo$ and the mapping $f\col\rat\to\topstwo$
defined by $f(a) := (a,\bot)$. Similarly, we can define $\TOPTWO$ over
the integers.

Note that $\SUM$, $\COUNT$ and $\PRTY$ are group aggregation 
functions,
while $\MAX$ and $\TOPTWO$ are idempotent. The aggregation function
$\PROD$  is also a group aggregation function, when defined 
over~$\ratnozero$.
However, one can prove that $\CNTD$, $\PROD$ (over $\rat$ or $\intg$)
and $\AVG$ 
are not monoid aggregation functions. 
\end{examples}


\section{Disjunctive Aggregate and Non-Aggregate Queries}
        \label{sec-conj:and:disj:queries}
We now introduce conjunctive and disjunctive queries with negated subgoals
and review their basic properties.
We use standard Datalog syntax extended by aggregation functions.

\subsection{Syntax of Non-aggregate Queries}
        \label{subsec-syntax:of:disjunctive:queries}
Predicate symbols are denoted as $p$, $q$ or $r$.
A~{\em term,\/} denoted as $s$ or $t$, is either a variable or a constant.
A {\em relational atom\/} has the form $p(\dd s k)$, where $p$ is a
predicate of arity $k$. We also use the notation $p(\tpl s)$,
where $\tpl s$ stands for a tuple of terms $(\dd s k)$.  
Similarly, $\tpl x$ stands for a tuple of variables.  
An {\em ordering atom\/} or {\em comparison\/} has the form 
$s_1 \osps{\rho} s_2$, where $\rho$ is one of the ordering predicates 
$<$, $\leq$, $>$, $\geq$, or $\neq$.
A relational atom can be {\em negated}. 
A relational atom that is not negated is {\em positive}.
A literal is a positive relational atom, a negated relational atom, or a 
comparison. 
A {\em condition,\/} denoted as $A$, is a conjunction of literals.
A condition $A$ is {\em
  safe\/}~\cite{Ullman-Principles:Of:DB:And:KB:Systems} if 
every variable appearing in $A$ either appears in a positive
relational atom or is equated with such a variable. Throughout this paper
we will assume that all conditions are safe.

A {\em query\/} is a non-recursive  expression of the form 
\begin{eqnarray*}
        q(\tpl x) \qif  A_1 \OR \cdots \OR A_n,
\end{eqnarray*}
where 
each $A_i$ is a condition containing all the variables appearing 
in the tuple $\tpl x$.
The variables that occur in the head, i.e., in $\tpl x$,  are
the {\em distinguished\/} variables of the query.
Those that occur only in the body are the {\em nondistinguished\/}
variables.

A query is {\em conjunctive\/} if it contains only one disjunct.
A query is {\em positive\/} if
it does not contain any negated relational atoms.
By abuse of notation, we will often refer to a query by its head 
$q(\tpl x)$ or simply by the predicate of its head $q$.

\subsection{Semantics of  Non-aggregate Queries}
        \label{ssec-query:semantics}

Databases are sets of
ground relational atoms and are denoted by the letter $\D$.
The carrier of $\D$, written $\carr \D$,
is the set of constants occurring in $\D$.
In this paper we assume that the constants in a database are either
integers or rational numbers.
We define how a query $q$, evaluated over a database $\D$, 
gives rise to a {set} of tuples $\DBD q$.

An {\em assignment\/} $\gamma$ for a condition $A$
is a mapping of the variables and constants appearing in $A$ to
constants, such that each constant is mapped to itself.
Assignments are naturally extended to tuples, atoms and other
complex syntactical objects.
For $\tpl s = (\dd s k)$ we let $\gamma(\tpl s)$ denote
the tuple $(\gamma(s_1),\ldots,\gamma(s_k))$.
The application of an assignment to other syntactical objects
is defined analogously.
{\em Satisfaction\/} of atoms and of conjunctions of atoms 
by an assignment \wrt a database or \wrt a semantic structure
are defined in the obvious way.
Sometimes, we will refer to assignments as {\em instantiations\/}
and to the result of applying an assignment to a syntactical object
as an {\em instantiation\/} of that object.

For the interpretation of comparisons it makes a difference whether
they range over a dense order, like the rational numbers, or a
discrete order, like the integers.  A conjunction of comparisons,
like $0 < y < z < 2$, 
may be satisfiable over the rational numbers, 
but not over the integers.

For a given database $\D$, a query $q(\tpl x) \qif A_1\OR\cdots\OR A_n$ 
defines a new relation 
\begin{eqnarray}
        \label{eqn-set:semantics}
   \DBD q  := 
     \bigcup_{i=1}^n
       \,\, \bigset{ \gamma(\tpl x) \bigmid
                 \mbox{$\gamma$ satisfies $A_i$ \wrt\ $\D$} } . 
\end{eqnarray}

Chaudhuri and Vardi~\cite{Chaudhuri:Vardi-Real:Conjunctive:Queries-PODS}
have introduced {\em bag-set semantics,\/} 
which records the multiplicity with which a tuple occurs as an answer
to the query.
The definition in~(\ref{eqn-set:semantics}) can be turned into one
for bag-set semantics by replacing set braces by multisets and set
union by multiset union.
{\em Bag semantics\/} 
\cite{Chaudhuri:Vardi-Real:Conjunctive:Queries-PODS}
differs from bag-set semantics in that both database relations and
relations created by queries are multisets of tuples.

\subsection{Syntax of Aggregate Queries}

In~\cite{Nutt:Et:Al-Equivalences:Among:Aggregate:Queries-PODS,%
        Cohen:Et:Al-Rewriting:Aggregate:Queries-PODS}
we have shown that equivalence of positive disjunctive 
queries with several aggregate terms can be reduced to
equivalence of queries with a single aggregate term.
Using a similar proof it is possible to show that this still holds if
the queries can contain negated subgoals. 
For this reason, we consider in the present paper only queries having
a single aggregate term in the head.
We give a formal definition of the syntax of such queries. 

An {\em aggregate term\/} is an expression built up using variables
and an aggregation function.
For example $\COUNT$ and $\SUM(y)$ are aggregate terms.
We use $\alpha(\tpl y)$ as an abstract notation for an aggregate term. 
Note that $\tpl y$ can be the empty tuple as in the case of the
functions $\COUNT$ or $\PRTY$.

An {\em aggregate query\/} is a query augmented by an aggregate
term in its head. Thus it has the form
\begin{eqnarray}
        \label{eqn-agg:query}
        q(\tpl x,\alpha(\tpl y))\qif A_1 \OR \cdots \OR A_n.
\end{eqnarray}
In addition, we require that
\begin{itemize}
\item no variable $x\in\tpl x$ occurs in $\tpl y$;
\item each condition $A_i$ contains all the variables in $\tpl x$ and in 
      $\tpl y$.
\end{itemize}
We call $\tpl x$ the {\em grouping variables\/} of the query. 
If the aggregate term in the head of a query has the form $\alpha(\tpl y)$,
we call the query an {\em $\alpha$-query\/}
(\eg, a $\MAX$-query).

\subsection{Semantics of Aggregate Queries}
        \label{subsec-sem:agg:query}

Consider an aggregate query $q$ as in Equation~\eqref{eqn-agg:query}.
We define how, for a database $\D$, 
the query yields a new relation $\DBD q$.
We proceed in two steps.

We denote the set of assignments $\gamma$ over $\D$ that satisfy one
of the disjuncts $A_i$ in the body of $q$ as $\ASSqD$.
We assume that such a $\gamma$ is defined 
only for the variables that occur in $A_i$.
Moreover, if an assignment $\gamma$ satisfies two or more disjuncts,
we want it to be included as many times in $\ASSqD$ as there are
disjuncts it satisfies.
To achieve this, we assume that there are as many copies of $\gamma$
in $\ASSqD$ as there are disjuncts that $\gamma$ satisfies, and that 
each copy carries a label indicating which disjunct it satisfies.%
\footnote{%
We could make this more formal by defining $\ASSqD$ to consist of
pairs $(\gamma,i)$, where $\gamma$ is an (ordinary) assignment and 
$i$ the index of a condition that it satisfies.
However, to avoid charging our notation with too much detail, we
prefer to introduce the concept of \quotes{labeled assignments}
informally.}

Recall that $\tpl x$ are the grouping variables of $q$ and $\tpl y$ are
the aggregation variables. 
For a tuple~$\tpl d$, let $\ASSqDd$ be the subset of $\ASSqD$
consisting of labeled assignments $\gamma$ with 
$\gamma(\tpl x) = \tpl d$.
In the sets $\ASSqDd$, we group those satisfying assignments
that agree on $\tpl x$.  
Therefore, we call $\ASSqDd$ the {\em group\/} of $\tpl d$.

Let $A$ be a set of labeled assignments and $\tpl y$ be a tuple of
variables for which the elements of $A$ are defined.
Then we define the {\em restriction of $A$\/} to $\tpl y$ as
the multiset
\begin{eqnarray*}
        \restr A{\tpl y} := 
           \bag{\gamma(\tpl y) \mid \gamma\in A}.
\end{eqnarray*}
We can apply the restriction operator to $\ASSqDd$.
If $\alpha(\tpl y)$ is an aggregate term, we can apply $\alpha$ to
the multiset $\restr A{\tpl y}$, which results in the aggregate
value $\alpha(\restr A{\tpl y})$.  
As an alternative notation, we define
\[
        \alpha(\tpl y)\downarrow A := \alpha(\restr A{\tpl y}).
\]
Now we define the result of evaluating $q(\tpl x,\alpha(\tpl y))$ 
over $\D$, denoted $\DBD q$, by
\begin{eqnarray*}
        \DBD q  :=  
        \bigset{ \big(\tpl d, \alpha(\tpl y) \downarrow \ASSqDd\big) \bigmid 
                \tpl d = \gamma(\tpl x) \mbox{ for some }\gamma\in\ASSqD}.
\end{eqnarray*}
Similarly as for non-aggregate queries, $\DBD q$ is a set of tuples.

\subsection{Equivalence}
        \label{subsec-equivalence}

Two queries $q$ and $q'$, aggregate or non-aggregate,
are {\em equivalent\/}, denoted $q\equiv q'$,
if over every database they return identical sets of results,
that is, if 
$\DBD q = \DBD{q'}$ for all databases $\D$.
For positive non-aggregate queries,
equivalence is decidable and
has been characterized in terms of the existence of 
query homomorphisms
\cite{Chandra:Merlin-Conjunctive:Queries-STOC,
        Sagiv:Yannakakis-Union:And:Difference-JACM,     
        Johnson:Klug-Optimizing:Conjunctive:Queries-SIAMJ}.
Levy and Sagiv have shown that equivalence is still decidable for
disjunctive queries with negated atoms
\cite{Levy:Sagiv-Sem:Query:Opt-PODS}.

In~\cite{Nutt:Et:Al-Equivalences:Among:Aggregate:Queries-PODS,%
         Cohen:Et:Al-Rewriting:Aggregate:Queries-PODS},
we have proved decidable characterizations for the equivalence of 
positive conjunctive and disjunctive aggregate queries with
the operators $\MAX$, $\COUNT$, and $\SUM$.
Note that two non-aggregate queries $q(\tpl x)$ and $q'(\tpl x)$ 
are equivalent under bag-set semantics if and only if 
the $\COUNT$-queries $q(\tpl x,\COUNT)$ and $q'(\tpl x,\COUNT)$
are equivalent.
Thus, characterizations of the equivalence of $\COUNT$-queries
immediately yield criteria for non-aggregate queries to be
equivalent under bag-set semantics.


\section{Bounded Equivalence}
        \label{sec-equivalence}
Our goal is to reduce the problem of deciding
equivalence of two aggregate queries over all possible
data\-bases to the problem of deciding {\em local equivalence,\/}
that is, equivalence over databases 
containing no more constants than the size of the queries.
In this section, we present the conditions necessary for the more general 
{\em bounded equivalence problem\/} to be decidable. 

Let $N$ be a nonnegative integer.
We say that two queries $q$ and $q'$ are {\em $N$-equivalent,\/} 
denoted $q \nequiv N q'$,
if for all databases $\D$ 
whose carrier has at most $N$ elements,
we have $\DBD q = \DBD{q'}$.
The {\em bounded equivalence problem\/} for a class of queries
is to decide, 
given $N > 0$ and queries $q$ and $q'$ from that class,
whether $q \nequiv N q'$.

Let $A$ be a condition. The {\em variable size\/} of $A$ is the number
of variables in $A$.
Let $q$ be a disjunctive query.
The {\em variable size\/} of $q$ is the maximum of the variable sizes
of the conditions in $q$. 
If a query contains an equality $y = z$, 
it does not matter for the proofs later on
whether the variables $y$ and $z$ are counted once or twice.

The {\em term size\/} of a query is the total number of constants
occurring in that query plus the variable size.
The {\em term size\/} of a pair of queries $q$ and $q'$ is the total
number of constants occurring in at least one of $q$ or $q'$ plus
the maximum of the variable sizes of $q$ and $q'$.
We denote 
the term size of $q$ as $\tsize q$ and 
the term size of $q$ and $q'$ as $\tsize{q,q'}$.
We say that two queries $q$ and $q'$ are {\em locally equivalent\/} 
if $\DBD q \nequiv{\tsize{q,q'}} \DBD{q'}$,
that is, if $q$ and $q'$ return identical results over all databases
with at most $\tsize{q,q'}$ constants.

Clearly, two queries are equivalent if and only if 
they are $N$-equivalent for all $N > 0$.
However, the decidability of bounded equivalence for a class of queries
does not necessarily imply that equivalence is decidable.
Sections~\ref{sec-decomposition} 
        and~\ref{sec-reduction} 
establish criteria for this implication to hold.
Moreover, decidability of $N$-equivalence, for a fixed $N$, does not 
imply decidability of local equivalence, since in the latter problem
the size of the databases to be tested depends on the size of the 
queries.

\begin{proposition}[Bounded and Local Equivalence]
If the bounded equivalence problem is decidable for a class of queries,
then local equivalence is decidable, too.
\end{proposition}

\begin{proof}
Deciding local equivalence of $q$ and $q'$ boils down to
deciding bounded equivalence of $q$ and $q'$ for $N = \tsize{q,q'}$.
\end{proof}

In the rest of this section we study the decidability of the 
bounded equivalence problem for several aggregation functions.
Note that $N$-equivalence is not necessarily a trivial property.
Even if the size of databases is bounded, there are still infinitely
many databases whose size is below the bound, and the aggregation
results may well depend on the values of the constants in the given database.

We introduce the notion of shiftable aggregation functions and of
order-decidable aggregation functions. 
We show that shiftable aggregation functions are a special case of
order-decidable aggregation functions.
Finally, we prove that
bounded equivalence is decidable exactly for queries with order-decidable
aggregation functions.

\subsection{Shiftable Aggregation Functions}
We introduce the notion of {\em shiftable aggregation functions}. 
Intuitively, the value of such a function does not depend on the
specific values in a multiset, but only on the ordering of the
elements.

Let $D$ and $D'$ be subsets of a domain $\I$ and 
$\phi\col D\to D'$ be a function.
We say that $\phi$ is a {\em shifting function\/} over $\I$
if for all $d$,~$d'\in D$ we have
\begin{eqnarray*}
        d < d'\  \Rightarrow \ \varphi(d) <  \varphi(d'). 
\end{eqnarray*}
In other words, a shifting function over a domain 
is a strictly monotonic function
from one subset of the domain to another subset.
A shifting function is applied to bags as one would expect.
Let $\alpha$ be an aggregation function that is defined over
$\I^k$. 
We say that $\alpha$ is {\em shiftable\/} if for all subsets $D$ and
$D'$ of $\I$,
for all shifting functions $\varphi\col D\to D'$,
and for all bags $B$ and $B'$ with elements in $D^k$, 
we have
\begin{eqnarray*}
        \alpha(B) = \alpha(B') 
        \iff 
        \alpha(\varphi(B)) = \alpha(\varphi(B')).
\end{eqnarray*}

\begin{proposition}[Shiftable Aggregation Functions]
        \label{prop:shiftable:functions} 
The aggregation functions 
 $\PRTY$, $\CNTD$, $\COUNT$, $\MAX$ and $\TOPTWO$
are shiftable. 
\end{proposition}

\begin{proof} 
The results of the aggregation functions $\PRTY$ and $\COUNT$ depend
only on the number of elements in the bag to which they are
applied. Applying a shifting function to a bag does not affect this
number. Therefore, these functions are shiftable. Similarly, the
result of the aggregation function $\CNTD$ depends only on the number
of distinct elements in the bag to which it is applied. Since shifting
functions are always injective, $\CNTD$
is also shiftable.

The aggregation function $\MAX$ chooses the greatest element in a bag. 
The order of the elements is preserved by a shifting function. 
Thus, $\MAX(\varphi(B)) = \varphi(\MAX(B))$. 
By definition,  $\varphi$ is an injection. 
Therefore, 
$\MAX(B) = \MAX(B')$ if and only if 
  $\varphi(\MAX(B)) = \varphi(\MAX(B'))$, 
which is true if and only if 
$\MAX(\varphi(B)) = \MAX(\varphi(B'))$. 
Hence, $\MAX$ is shiftable.

Using similar reasoning to that of $\MAX$ it is easy to see 
that $\TOPTWO$ is shiftable. 
\end{proof}

Note, however, that the aggregation functions $\SUM$ and $\PROD$ 
are not shiftable.
For example, consider the bags 
$B = \bag{2,\,2}$ and $B' = \bag{4}$ and 
suppose $\varphi$ is a shifting function  with 
$\varphi(2) = 3$ and $\varphi(4) = 5$.
Then $\SUM(B) = \SUM(B') = \PROD(B) = \PROD(B') = 4$, while 
neither $\SUM$ nor $\PROD$ agree on $\varphi(B) = \bag{3,\,3}$
and $\varphi(B') = \bag{5}$.

\subsection{Order-Decidable Aggregation Functions}\label{sec:order-decid-aggr}
Before defining order-decidable aggregation functions, we present some
auxiliary definitions.
Given a domain $\I$, a conjunction of ordering atoms $L$, and an
ordering atom $t \osps{\rho} t'$, 
we define in the standard way when
$L$ entails $t \osps{\rho} t'$ \wrt $\I$, denoted $L\dommodels \I
t \osps{\rho} t'$, and when  $L$ is satisfiable \wrt $\I$.

We say that $L$ is a {\em complete ordering\/} of a set of terms $T$
{\em with respect to $\I$\/} if for every two terms
$t$,~$t'\in T$, exactly one of the following holds:
\begin{itemize}
\item $L\dommodels \I t < t'$;
\item $L\dommodels \I t > t'$;
\item $L\dommodels \I t = t'$.
\end{itemize}
Note that by definition, complete orderings are satisfiable.


Let $\alpha$ be an aggregation function over $\I^k$.
An {\em ordered identity\/} for $\alpha$ is a formula 
\begin{equation}
  \label{e:decidable}
   L \to \alpha(B) = \alpha(B')
%
\end{equation}
where $L$ is a complete ordering of some set of terms $T$ \wrt\ $\I$,
and $B$ and $B'$ are bags containing $k$-tuples of terms from $T$.
We say that $\alpha$ is {\em order-decidable\/} over $\I$
if the validity of ordered identities for $\alpha$ 
is decidable over $\I$.
Note that the validity of an ordered identity 
may be dependent on~$\I$.

Formula~\eqref{e:decidable} is valid if 
for every assignment $\delta$ that maps the variables in $L$ to $\I$
and satisfies $L$,
we have that $\alpha$ yields the same values
when applied to $\delta(B)$ and to $\delta(B')$.

\eat{
Formula~\eqref{e:decidable} makes a statement 
about all assignments $\delta$ 
that map the variables in $B$ and $B'$ to elements of $\I^k$.
It says that if $\delta$ satisfies $L$, 
then the values of $\alpha$, 
when applied to $\delta(B)$ and to $\delta(B')$,
are equal.
}

\begin{example}
It is easy to see that the function $\CNTD$ is
order-decidable over any domain. 
Consider, for example, 
the bags $B = \bag{1,\, 2,\, u}$ and $B' = \bag{v,\, v,\, 7,\, 8}$ 
and an arbitrary complete ordering~$L$ of 
$\set{1,\, 2,\, u,\, v,\, 7,\, 8}$. 
It is straightforward to decide whether the formula 
\[ 
     L \to \CNTD(\bag{1,2,u}) = \CNTD(\bag{v,v,7,8})
\]
is valid, since for any assignment $\delta$ satisfying $L$
the number of distinct values in the bags 
$\delta(B)$ and~$\delta(B')$
is not dependent on the values assigned to $u$ and $v$.
In fact, the number of distinct elements that are contained in 
$\delta(B)$ and $\delta(B')$
depends entirely on the ordering $L$.
\end{example}

It is not by chance that the function $\CNTD$ is order-decidable 
over all domains.
It is actually a consequence of the fact that $\CNTD$ 
is a shiftable aggregation function.

\begin{theorem}[Shiftable Implies Order-Decidable]
 \label{p:shiftable:universal}
Let $\alpha$ be a shiftable aggregation function 
defined over  $\I^k$.
Then $\alpha$ is order-decidable over $\I$.
\end{theorem}

\begin{proof}
Let $L \to \alpha(B) = \alpha(B')$ be an ordered identity
as in \eqref{e:decidable}.
In principle, to check this identity for validity,
one has to verify that for all $\delta$ satisfying $L$
the equality $\alpha(\delta(B)) = \alpha(\delta(B'))$ holds.
We will show that it is sufficient to verify the equality for a single
$\delta$ if $\alpha$ is shiftable.

Suppose that $\alpha$ is a shiftable aggregation function over $\I^k$.
Let $\delta\col T\to \I$ be an assignment.
Clearly, if $\delta$ satisfies $L$,
then the following conditions hold:
\begin{itemize}
 \item  $\delta$ maps all constants to themselves;
 \item  for all $t$, $t'\in T$ and all ordering
        predicates $\rho$ we have that
        \[ 
            t \osps{\rho} t' \in L 
            \ \ \impl \ \ 
            \dommodels I \delta(t)\osps{\rho}\delta(t') . 
        \]
\end{itemize}

Consider Formula~\eqref{e:decidable}.  
Since $L$ is a complete ordering, $L$ is satisfiable \wrt $\I$. 
Let $\delta$ be an assignment satisfying $L$.
Now, let $\delta'\col T\to \I$ be a second assignment that satisfies $L$.  
We assume without loss of generality that there are no two different
terms $t_1$, $t_2 \in T$ for which $L \dommodels \I t_1 = t_2$.  
(If there were such terms we could remove one of them by renaming.)
Hence, $\delta$ and $\delta'$ are injections. 
Thus, the function $\delta' \circ \inv\delta$ is well defined.

Since both $\delta$ and $\delta'$ preserve order, 
$\delta' \circ \inv\delta$ is a shifting function.
Thus, $\alpha(\delta(B)) = \alpha(\delta(B'))$ implies 
$\alpha(\delta'(B)) = \alpha(\delta'(B'))$, as required.
\end{proof}

The other direction of Theorem~\ref{p:shiftable:universal} does
not hold. 
An aggregation function can be order-decidable over a given
domain even if it is not shiftable. 
For example, the aggregation functions $\SUM$ and $\AVG$ are 
order-decidable, although they are not shiftable.

\begin{proposition}[Order Decidability of Sum and Average]
       \label{prop:sum:avg:order-decidable} 
The aggregation functions $\SUM$ and $\AVG$  are
order-decidable over $\intg$ and over $\rat$.
\end{proposition}

\begin{proof} 
For the aggregation function $\SUM$, 
 Formula~\eqref{e:decidable}
 can be expressed using Presburger arithmetic. 
Recall that Presburger arithmetic is the first-order theory of addition.
Presburger showed~\cite{Presburger} that 
Presburger integer arithmetic 
(i.e., where the variables range over the integers) 
is decidable. 
Similarly, Presburger rational arithmetic is also known to be
decidable~\cite{Kreisel-Krivine:Model-Theory}. 
Therefore, $\SUM$ is order-decidable.

The order-decidability of $\AVG$ follows in a straightforward fashion
from the order-decidability  of  $\SUM$, as we now show. 
Let $B$ be a bag of size $N$.
We use $N\otimes B$ to denote the bag derived from $B$ by
increasing the multiplicity of each term in $B$ by a factor of $N$.
Thus, $N\otimes B$  contains exactly the same
terms as those in $B$. If a term $t$ appears in $B$ exactly $k$ times,
then $t$ appears in $N\otimes B$ exactly $Nk$ times.

Consider bags of numbers $B$ and $B'$. Suppose that $B$ is of size $N$
and $B'$ is of size $N'$.
Observe that
\begin{align*}
 \AVG(B) = \AVG(B')  & {} \iff N'\,\SUM(B) = N\,\SUM(B') 
\iff \SUM(N'\otimes B) = \SUM(N\otimes B')
\end{align*}
Therefore, 
\[
   L \to \AVG(B) = \AVG(B')
\]
is valid if and only if 
\[
   L \to \SUM(N'\otimes B)=\SUM(N\otimes B')
\]
is valid, where $N$ and $N'$ are the cardinalities of $B$ and $B'$,
respectively. Hence, since $\SUM$ is order-decidable, $\AVG$ is
also order-decidable.
\end{proof}

The aggregation function $\PROD$ is also order-decidable. In order to
show this result we first present a few necessary definitions and
lemmas. These are needed when considering $\PROD$ over the integers.

Let $T$ be a set of terms and let $L$ be a complete
ordering of $T$.
We say that $T$ is {\em reduced\/} with respect to $L$ 
and to a domain $\I$ if 
\begin{itemize}
\item there are no different variables $x$ and $y$ occurring in $T$ 
      such that $L \dommodels \I x = y$;
\item there is no variable $x$ occurring in $T$ 
      and no constant $d$ in $\I$ such that $L \dommodels \I x = d$.
\end{itemize}
%
%
We say that a constant $c$ is a {\em possible value\/} 
for a variable $x\in T$ with respect to $L$ and $\I$ 
if there is an assignment for the variables in $T$ 
with constants from $\I$ that satisfies 
$L$ and maps $x$ to the value $c$. 
Observe that if $T$ is reduced with respect to $L$ and $\I$, 
then there are at least two different possible values 
for each variable in $T$.
Also, note that $T$ may be reduced \wrt $L$ over the rational numbers, 
but not over the integers.
For instance, $T = \set{0,\, x,\, 2}$ is reduced \wrt
$L = \set{0<x,\, x<2}$ over the rational numbers, 
but over the integers, $L$ entails that $x=1$.

\begin{lemma}[Assignments for Possible Values]
  \label{lemma:two:instantiations} 
Let $L$ be a complete ordering of the terms in $T$. 
Suppose that $T$ is reduced with respect to $L$ and to $\I$. 
Let $x$ be a variable in $T$ and 
let $c_1$ and $c_2$ be possible values for $x$ 
with respect to $L$ and~$\I$.  
Then there are 
assignments $\delta_1$ and $\delta_2$ for the terms in $T$ 
that satisfy $L$, 
are equal on all terms other than $x$, 
and such that 
$\delta_1(x) = c_1$ and $\delta_2(x) = c_2$.
\end{lemma}

\begin{proof}
Since $c_1$ and $c_2$ are possible values for $x$, there are
assignments $\delta_1$ and $\delta_2$ that satisfy $L$
such that $\delta_i(x) = c_i$, for $i = 1$,~$2$.
Note that $\delta_1$ and $\delta_2$ may also
differ on additional terms in $T$. 

Let $\delta$ be the assignment for the terms in $T$ defined by
\[ 
  \delta(t) := 
   \begin{cases}
     \MIN\set{\delta_1(t),\, \delta_2(t)} 
              & \mbox{if $L\dommodels \I t \leq x$} \\
     \MAX\set{\delta_1(t),\, \delta_2(t)} 
              & \mbox{if $L\dommodels \I t > x$.} 
   \end{cases}
\]
We show that $\delta$ satisfies $L$. Let $t$ and $t'$ be terms in
$T$. Suppose that $L\dommodels \I t < t'$. We consider two cases. 

\CASE{1}  
Suppose that $L\dommodels \I t'\leq x$. 
Let $i$ be such that
$\delta(t') = \delta_i(t')$. 
Then, $\delta(t) \leq \delta_i(t) < \delta_i(t') =
\delta(t')$. Therefore, $\delta$ satisfies $t<t'$.

\CASE{2}
Suppose that $L\dommodels \I x < t'$. 
Let $i$ be such that $\delta(t) = \delta_i(t)$. 
Then, $\delta(t) = \delta_i(t) < \delta_i(t') \leq \delta(t')$.
Therefore, $\delta$ satisfies $t<t'$.

Since $L$ is a complete ordering and $t < t'$ was arbitrary,  
it follows that $\delta$ satisfies $L$.
In a similar fashion we define the assignment $\delta'$ as
\[ 
  \delta'(t) := 
   \begin{cases}
     \MIN\set{\delta_1(t),\, \delta_2(t)} 
              & \mbox{if $L\dommodels \I t < x$} \\
     \MAX\set{\delta_1(t),\, \delta_2(t)} 
              & \mbox{if $L\dommodels \I t \geq x$} 
   \end{cases}
\]
We can show, as above, that $\delta'$ satisfies $L$. Clearly $\delta$
and $\delta'$ are equal on all terms other than $x$. One of the
assignments $\delta$ or $\delta'$ maps $x$ to $c_1$ and one maps $x$
to $c_2$.  Therefore, we have found assignments as required.
\end{proof}

\begin{proposition}[Order Decidability of Product]
  \label{prop:prod:order-decidable} 
The aggregation function $\PROD$ is order-decidable over $\intg$ and
over $\rat$. 
\end{proposition}

\begin{proof}
Let $T = \set{\dd tn}$ be a set of terms 
with constants from $\intg$ or $\rat$, and 
let $L$ be a complete ordering of $T$. 
Let $B$ and $B'$ be bags of terms from $T$. 
We will show that it is possible to decide whether 
\begin{align}
        \label{eqn-order:decidability:prod}
   L \to \PROD(B) = \PROD(B')
\end{align}
is valid over $\intg$ or $\rat$, respectively.
Clearly, Formula~\eqref{eqn-order:decidability:prod} 
is valid if $\PROD(\delta(B)) = \PROD(\delta(B'))$ 
for all assignments $\delta$ that satisfy $L$. 

There may be assignments that satisfy $L$ and 
map variables to the constant 0. 
It is important to be able to recognize these assignments. 
Let $T' = T \cup \set{0}$. 
Let $L'$ be a complete ordering of $T'$
that is a conservative extension of $L$.
Formally, this means that for all terms $t$, $t'\in T$, 
the orderings $L$ and $L'$ imply 
the same relationship between $t$ and $t'$. 
Note that if $0\in T$, then $L'$ must be equivalent to $L$. 
There are only finitely many conservative extensions 
$L'$ of $L$,
and an assignment satisfies $L$ if and only if 
it satisfies one of the extensions $L'$.
Thus, to prove our claim, we can assume without loss of generality
that $L$ in Formula~\eqref{eqn-order:decidability:prod}
is a complete ordering of a set of terms that contains the constant 0.

Furthermore, we can assume without loss of generality, 
that in Formula~\eqref{eqn-order:decidability:prod}
the set of terms $T$ is reduced \wrt $L$.
Otherwise, 
whenever $T$ contains a variable $y$ and a term $t$
such that $y$ and $t$ are distinct, but $L\dommodels \I y = t$,
then we replace $y$ with $t$ 
for every occurrence of $y$ in $L$, $B$ and $B'$.
Eventually, we end up with a set of terms $\bar T$,
a complete ordering $\bar L$ of $\bar T$, 
and bags $\bar B$, $\bar B'$ of terms from $\bar T$
such that $\bar T$ is reduced \wrt $\bar L$, and
$\bar L \to \PROD(\bar B) = \PROD(\bar B')$
is valid if and only if 
Formula~\eqref{eqn-order:decidability:prod} is valid.

Next, we rewrite the equation \quotes{$\PROD(B) = \PROD(B')$}.
We note that for every assignment, 
$\PROD(B)$ yields the same value as the polynomial
$c\, u_1^{m_1} \cdots  u_k^{m_k}$, 
where 
\begin{itemize}
 \item $c$ is the product of all the constants in $B$;
 \item $\dd u k$ are all the variables in $T$;
 \item $m_i$ is the multiplicity of $u_i$ in $B$.
\end{itemize}
Similarly, $\PROD(B')$ yields the same value as some polynomial
$d\, u_1^{n_1} \cdots u_k^{n_k}$.
Now, deciding the validity of Formula~\eqref{eqn-order:decidability:prod}
amounts to deciding whether the equation 
\begin{align}
   \label{eqn-prod:equality}
  c \, \big(\delta(u_1)\big)^{m_1}\cdots 
              \big(\delta(u_k)\big)^{m_k} 
= d \, \big(\delta (u_1)\big)^{m_1}\cdots 
              \big(\delta (u_k)\big)^{m_k}
\end{align}
holds for all assignments $\delta$ satisfying $L$.

If $c = d = 0$, then clearly 
Equation~\eqref{eqn-prod:equality} holds for any $\delta$. 
Similarly, Equation~\eqref{eqn-prod:equality} holds
for any $\delta$ if $c = d$ and $m_i = n_i$ for all $i$.
We show that if neither of the above conditions holds, 
then there is an assignment $\delta$ that satisfies $L$
and for which Equation~\eqref{eqn-prod:equality} is not true. 
We consider two cases.

\CASE{1}
  Suppose that $m_i = n_i$ for all $i$, 
  however $c\neq d$. 
  Since $L$ is a complete ordering, it is satisfiable. 
  Let $\delta$ be an assignment that satisfies $L$. 
  Since the set of terms $T$ is reduced \wrt $L$, 
  and $0$ is an element of $T$,
  the ordering $L$ imposes a strict inequality 
  between 0 and each variable. 
  Therefore,  $\delta$ cannot map any variable to the constant 0. 
  Hence, $\delta$ is a counterexample to the correctness of
  Equation~\eqref{eqn-prod:equality}.

\CASE{2}
  Suppose that one of $c$ or $d$ is non-zero and 
  that there is an index $i$ such that $m_i\neq n_i$.  
  Again, since $T$ is reduced \wrt $L$,
  there are at least two possible values for $u_i$, 
  say~$c_1$ and~$c_2$. 
  Let $\delta_1$ and $\delta_2$ be assignments 
  that agree on all terms other than $u_i$ and 
  that satisfy $\delta_j(u_i) = c_j$ for $j = 1$,~$2$. 
  Such assignments exist according to
  Lemma~\ref{lemma:two:instantiations}. 

  As before, $\delta_1$ and $\delta_2$ cannot map 
  any variable to the constant 0. 
  If $c = 0$ and $d \neq 0$, then both  $\delta_1$ and $\delta_2$
  are counterexamples to the correctness of
  Equation~\eqref{eqn-prod:equality}. 
  Similarly, they are counterexamples if $c \neq 0$ and $d = 0$. 
  Therefore, assume that $c\neq 0$ and $d \neq 0$.

  Suppose, by way of contradiction, that
  Equation~\eqref{eqn-prod:equality} holds 
  for both $\delta_1$ and $\delta_2$. 
  Then, 
  \begin{align*}
     c \, \big(\delta_j(u_1)\big)^{m_1}\cdots 
                 \big(\delta_j(u_k)\big)^{m_k} 
   = d \, \big(\delta_j (u_1)\big)^{m_1}\cdots 
                 \big(\delta_j (u_k)\big)^{m_k}
  \end{align*}
  for $j = 1, 2$.
  Since the assignment $\delta_2$ does not map any variable to 0, 
  we can divide the equation with $j = 1$ by the equation with $j= 2$. 
  Note that $\delta_1$ and $\delta_2$ are equal 
  on all terms other than $u_i$. 
  Therefore, after simplifying, we derive
  \begin{align}
      \label{frac:contra}
  \left(\frac{\delta_1(u_i)}{\delta_2(u_i)}\right)^{m_i} = 
    \left(\frac{\delta_1(u_i)}{\delta_2(u_i)}\right)^{n_i}.
  \end{align}
  Note that $\delta_1(u_i)/{\delta_2(u_i)}\neq 0$,
  since $\delta_1$ does not map any variable to the constant 0. 
  In addition, 
  ${\delta_1(u_i)}/{\delta_2(u_i)}\neq 1$, 
  since $\delta_1$ and $\delta_2$ differ on $u_i$. 

  Finally, ${\delta_1(u_i)}/{\delta_2(u_i)}\neq -1$, 
  since $\delta_1$ and $\delta_2$ must both map $u_i$ 
  to positive numbers or both map $u_i$ to negative numbers. 
  Therefore, Equation~\eqref{frac:contra} cannot
  hold and either $\delta_1$ or $\delta_2$ is a counterexample to
  the correctness of Equation~\eqref{eqn-prod:equality}. 
  This completes Case~2.

Thus, we have shown how to decide the validity of 
Formula~\eqref{eqn-order:decidability:prod}
over both, the integers and the rational numbers.
This completes the proof.
\end{proof}

\subsection{Decidability of Bounded Equivalence}
It is possible to show that 
bounded equivalence can be decided for $\alpha$-queries containing
comparisons that range over $\I$
if the aggregation function $\alpha$ is order-decidable over $\I$.
Actually,
bounded equivalence for $\alpha$-queries ranging over $\I$ 
is decidable {\em if and only if\/} 
$\alpha$ is order-decidable over $\I$. 
This gives a complete characterization of decidability of bounded
equivalence of aggregate queries with negation, disjunction, constants
and comparisons.
In addition, we derive as a direct result 
that bounded equivalence is decidable 
for queries with a wide range of common aggregation functions. 

\begin{theorem}[Bounded Equivalence and Order-Decidability]
    \label{t:bounded:equivalence}
Let $\alpha$ be an aggregation function over $\I^k$. 
Then the bounded equivalence problem is decidable 
for disjunctive $\alpha$-queries with comparisons ranging over $\I$ 
if and only if $\alpha$ is order-decidable over $\I$. 
\end{theorem}

\begin{proof}
\If 
Suppose that $\alpha$ is order-decidable over $\I$. 
Consider $\alpha$-queries $q$ and $q'$. 
We show how to check, given some $N > 0$, 
whether $q\nequiv N q'$.

Let $C$ be the set of constants appearing in $q$ or $q'$ 
and let $U$ be a set of $N$ variables. 
We use $T$ to denote $C\cup U$.
Let $P$ be the set of predicates appearing either in $q$ or in $q'$. 
The set $P$ contains predicates that appear either positively or 
negatively in the queries. 
We use $\arity(p)$ to denote the arity of a predicate $p\in P$. 
We denote by $\BASE$ the set of all atoms that can be created using
the terms in $T$ and the predicates in $P$. Formally,
\begin{align*}
  \BASE := \set{\,p(t_1,\ldots,t_{\scriptarity(p)}) \mid 
                       p\in P \mbox{ and } 
                       t_1,\ldots,t_{\scriptarity(p)}\in T}.
\end{align*}

\eat{
\begin{align*}
  \BASE := \set{\,p(t_1,\ldots,t_{\scriptarity(p)}) \mid {}
                       & p\in P \mbox{ and } \\
                       & t_1,\ldots,t_{\scriptarity(p)}\in T}.
\end{align*}

\begin{multline*}
\BASE := \\
        \set{\, p(t_1,\ldots,t_{\scriptarity(p)}) \mid p\in P 
                \mbox{ and }
                t_1,\ldots,t_{\scriptarity(p)}\in T \,}.
\end{multline*}
}

If $\delta$ is an assignment 
that maps variables in $U$ to elements of $\I$, 
and if $S$ is a subset of $\BASE$,
then instantiating $S$ by $\delta$ 
results in a database $\delta(S)$,
the carrier of which has at most $N$ elements.
To decide whether $q\nequiv N q'$ 
it is sufficient to evaluate the queries over databases of the 
form $\delta(S)$ where $S\incl\BASE$.
Essentially, if we consider only databases of this form,
we  rule out databases containing predicates 
not appearing in $q$ or $q'$. 
Clearly, such predicates cannot affect the evaluation of 
$q$ and $q'$. 

Consider now a fixed subset $S \incl \BASE$.
We will show how to check whether $q$ and $q'$ 
return the same results over all instantiations $\delta(S)$.
Since there exist infinitely many instantiations,
we cannot check each of them separately.
Instead, we divide the instantiations into finitely many equivalence 
classes over which we can decide the equivalence of $q$ and $q'$. 
The equivalence classes are defined 
by the complete orderings $L$ of $T$,
that is, for each $L$ we simultaneously check 
all instantiations $\delta(S)$ where $\delta$ satisfies $L$. 

In addition to $S$, consider a complete ordering $L$ of $T$. 
Instead of an instantiation $\delta(S)$,
we attempt to evaluate $q$ and $q'$ immediately over $S$, 
based on the ordering of terms defined by $L$.
We view the set $S$ equipped with this ordering as a database,
denoted $S_L$.
Obviously, given $S$ and $L$, 
we can compute the bags containing the tuples returned by $q$ and $q'$.
However, it is impossible 
to compute the values of the aggregation function $\alpha$ for these bags 
because $T$, and therefore the bags, may contain variables.
At this point, we make use of the fact that 
$\alpha$ is order-decidable over $\I$.

Suppose that the tuple of grouping variables $\tpl x$ 
in $q(\tpl x, \alpha(\tpl y))$ and $q(\tpl x, \alpha(\tpl y'))$ 
has length $k$.  
Let $\tpl t$ be a $k$-tuple of terms in $T$. 
Note that there are only finitely many such tuples 
because $T$ is finite.
Recall that $\ASStpl{q}{S_L}{\tpl t}$ is the set of assignments 
$\gamma$ that satisfy $q$ over $S_L$ and 
where $\gamma(\tpl x) = \tpl t$.
Consider the bag $B_{\tpl t}$ defined as 
\[
  B_{\tpl t} := \restr{\ASStpl{q}{S_L}{\tpl t}}{\tpl y},
\]
that is, $B_{\tpl t}$ consists of the restrictions of elements of 
$\ASStpl{q}{S_L}{\tpl t}$ to the variables in $\tpl y$.
Let $B'_{\tpl t}$ be defined analogously for $q'$.

Now, assume that there is an assignment $\delta$ that satisfies $L$
such that over the database $\delta(S)$ 
the queries $q$ and $q'$ do not return the same aggregate value 
for $\delta(\tpl t)$.
This is the case if and only if the formula
\[ 
           L \to \alpha(B_{\tpl t})=\alpha(B'_{\tpl t})
\]
is not valid over $\I$.
Since $\alpha$ is order-decidable over $\I$, 
the validity of the formula can be determined.

\medskip
\OnlyIf
We show that if bounded equivalence of 
$\alpha$-queries ranging over $\I$ is decidable, 
then $\alpha$ is order-decidable over $\I$. 
To simplify our notation, we assume without loss of generality 
that $\alpha$ is a unary function.
Let $T=\set{\dd tN}$ be a set of terms,
$L$ be a complete ordering of $T$, and
$B$ and $B'$ be bags of terms from $T$.
We will construct $\alpha$-queries $q$ and $q'$ such that 
\begin{equation}
 \label{eqn-bag:aggregates:proof}
   L \to \alpha(B) = \alpha(B')
%
\end{equation}
is valid over $\I$ if and only if $q\nequiv N q'$.

We assume without loss of generality 
that $L$ does not equate two different terms.
Otherwise, we could remove one of them by renaming. 
We define the condition $A$ as
\[ 
   A:= p(t_1) \AND p(t_2) \AND \ldots \AND p(t_N).  
\]
Suppose that $B = \bag{s_1,\ldots,s_m}$ and 
             $B'= \bag{s'_1,\ldots,s'_n}$.
We assume that $y$ is a {\em new\/} variable, 
i.e., $y$ does not appear in $B$ or $B'$. 
We define the conditions 
\begin{align*}
    A_i  & := A \AND L \AND y = s_i, \hspace{-10em}  & i & =1,\ldots, m  \\
    A'_j & := A \AND L \AND y = s'_j, \hspace{-10em} & j & =1,\ldots, n. 
\end{align*}
and we define the queries $q$ and $q'$ by
\[
     q(\alpha(y)) \qif \bigvee_{i=1}^{m} A_i    
         \hspace{2em}\mbox{and}\hspace{2em}
    q'(\alpha(y)) \qif \bigvee_{j=1}^{n} A'_j.
\]

Suppose that $q\nequiv N q'$. 
We prove this implies that 
Formula~\eqref{eqn-bag:aggregates:proof} is valid.
Let $\delta$ be an arbitrary assignment that satisfies $L$. 
We show that 
$\alpha(\delta(B)) = \alpha(\delta(B'))$. 
Consider the database $\D$ 
obtained by instantiating $A$ with $\delta$, i.e., 
\[ 
   \D := \set{ p(\delta (t_1)), \ldots, p(\delta (t_N)) }\,. 
\]
Clearly, $q$ and $q'$ retrieve $\alpha(\delta(B))$ 
and $\alpha(\delta(B'))$, respectively, over $\D$. 
The database $\D$ contains at most $N$ constants.
Thus, since $q \nequiv N q'$, the two aggregates are equal.

Now suppose that Formula~\eqref{eqn-bag:aggregates:proof} is valid.
We prove this implies that $q\nequiv N q'$. 
Let $\D$ be an arbitrary database containing at most $N$ constants. 
If $\D$ contains less than $N$ values, 
then no assignment over $\D$ can satisfy $L$, 
because $L$ is a complete ordering that 
does not equate variables.
Hence, $q$ and $q'$ will not return any value over~$\D$.

Therefore, assume that $\D$ contains exactly $N$ values. 
It is easy to see that $q$ is satisfiable over $\D$ 
if and only if $q'$ is satisfiable over $\D$. 
In such a case,
for each condition $A_i$ or $A'_j$ there is
exactly one satisfying assignment over $\D$,
say, $\gamma_i$ or $\gamma'_j$, respectively.
Moreover, the assignments $\gamma_i$ and $\gamma'_j$ 
agree on all variables except for $y$.
That is, there is an assignment $\delta$ for the variables in $T$
such that $\gamma_i$ and $\gamma'_j$ agree with $\delta$ on $T$,
for all $i$ and $j$.
Due to the definition of $A_i$ and $A'_j$,
we also have 
$\gamma_i(y) = \delta(s_i)$ and $\gamma'_j(y) = \delta(s'_i)$.
Thus, $q$ collects over $\D$ the bag 
$\bag{\gamma_1(y),\ldots,\gamma_m(y)} =
 \bag{\delta(s_1),\ldots,\delta(s_m)} = \delta(B)$
and returns the aggregate $\alpha(\delta(B))$.
Similarly, $q'$ returns the aggregate $\alpha(\delta(B'))$. 
From the assumption that
Formula~\eqref{eqn-bag:aggregates:proof} is valid,
and the fact that $\delta$ satisfies $L$, 
it follows that $\alpha(\delta(B)) = \alpha(\delta(B'))$. 
Hence, $q$ and $q'$ return the same values over $\D$. 
Since $\D$ was chosen arbitrarily, this proves that $q\nequiv N q'$. 
\end{proof}


\medskip
From the proof of Theorem~\ref{t:bounded:equivalence} we can derive an
upper bound on the complexity of determining whether $q \nequiv N q'$.
In fact, the proof describes a procedure for checking $N$-equivalence.
Suppose that there are $C$ constants in $q$ and $q'$. 
Let $T := C + N$. 
As the first step of the procedure, the set $\BASE$ is created.
This set contains all possible instantiations of the atoms in $q$ 
and $q'$ with the $C$ constants and with $N$ variables.
Clearly, the cardinality of this set is exponential in $T = C + N$. 
Then, each subset $S$ of $\BASE$ is considered. 
The number of such subsets is, thus, double exponential in $T$. 
A subset $S$ is considered in conjunction with a complete ordering $L$
of the $T$ terms.
Note that there are at most $2^{T-1}\, T!$ complete orderings of 
$T$ terms. 
(This is a rough upper bound, since we can arrange the $T$ terms in 
$T!$ orders and then place a \quotes{$<$} or \quotes{$=$} sign 
between each pair.) 
Thus, considering all complete orderings does not affect the already 
double exponential order of complexity.

For each pair $S$, $L$ we evaluate $q$ and $q'$. 
Evaluating $q$ roughly takes time $T^{|q|}$, 
where $|q|$ is the size of $q$,
since we must try all instantiations of the terms in $q$.
This too does not affect the order of complexity, since the computation
time is only exponential, while there are a double exponential number
of subsets $S$.
For each tuple $\tpl t$ that instantiates the grouping variables
and thus defines a group,
we check the validity of the ordered identity defined in
Formula~\eqref{e:decidable}  
for $L$ and the bags created. 

Each bag can have at most $T^{|q|}$ many elements.
However, the number of different elements is bounded by $T^k$, where
$k$ is the arity of the aggregation function $\alpha$.
Therefore, to represent a bag we only need space polynomial in $T$.
Thus, the size of the ordered identities to be checked for validity is
polynomial in $T$.
As long as this check takes no more than double exponential time,
the overall complexity is at most double exponential.
In many cases, this step is much more efficient.
For example, given the aggregation function $\COUNT$, this step only
requires checking the cardinality of the bags, and hence, is linear.

To summarize, if the validity of ordered identities of the
form~\eqref{e:decidable} can be checked in double exponential time, we
derive a double exponential upper bound for the complexity of checking
$N$-equivalence.

\medskip

\eat{ 
We derive a complexity bound from the proof of
Theorem~\ref{t:bounded:equivalence}. %

\eat{
Let $q$ and $q'$ be
$\alpha$-queries. We use $C(q,q')$ to denote the number of constants
in $q$ or $q'$. Similarly, $P(q,q')$ denotes the number of predicates
in $q$ or $q'$ and $A(q,q')$ denotes the maximum arity of any
predicate in $q$ or $q'$. 
We use $X(q)$ and $Y(q)$ to denote the number of
grouping and aggregation variables, respectively, in $q$. We will
sometimes write $A$ Suppose that
for given $L$, $B$ 
and $B'$, it is possible to determine the validity of
Formula~\eqref{e:decidable} in time $f_\alpha(|L|,|B|,|B'|)$. 
}

Let $C$ be the number of 
constants in either $q$ or $q'$. Let $X$ and $Y$ be the number of grouping
variables and aggregation variables, respectively, in 
$q$ and $q'$. Suppose that there are $P$ different predicates in $q$ and $q'$
and the maximum arity of any predicate is $A$. We use $Q$ to denote
the total size of $q$ and $q'$.
Suppose that for given
$L$, $B$ 
and $B'$, it is possible to determine the validity of
Formula~\eqref{e:decidable} in time $f_\alpha(|L|,|B|,|B'|)$.

\newcommand{\numLinExpansion}[1]{??}
\begin{theorem}[Complexity of Bounded Equivalence]
Let $q$ and $q'$ be alpha queries. 
Then it is possible to determine if $q\nequiv q'$ in time
\[  2\, f_\alpha(T,T^Y\!,T^Y)\,T^{X+N}\,2^{P{T}^A}\, \numLinExpansion T  \]
where $T = N+C$.
\end{theorem}

\begin{proof}
This follows from a careful analysis of the procedure for checking
bounded equivalence described in the proof of
Theorem~\ref{t:bounded:equivalence}. We first must construct the set
$\BASE$. This set contains, for each predicate, all possible atoms
that can be created using a total of $T$ terms. Thus, $\BASE$ has size
 $PT^A$. Then, for each subset $S$ of $\BASE$, of which there are
 $2^{P{T}^A}$, we check whether $q$ and $q'$ return the same values
 over all instantiations of $S$. 

For this purpose, we consider all
 complete orderings $L$ of the $T$ terms. A complete ordering divides
 the terms of $T$ into equivalence classes (of equal terms) and then
 determines an ordering among these equivalence classes. Thus, there
 are at most  \[ \numLinExpansion T \]
different complete orderings of $T$ terms. 

For each complete ordering $L$ of $S$, we consider all $T^X$ possible grouping
tuples $\tpl t$ that $\tpl x$ can be assigned to. 
Then, for each $\tpl t$, we compute $q$ and $q'$ over $S$ with ordering
$L$. This requires checking up to $T$ assignments of the variables of
$q$ and $q'$. 
\end{proof}

}

The following corollary
follows directly from
Theorems~\ref{p:shiftable:universal} and~\ref{t:bounded:equivalence}.

\begin{corollary}[Bounded Equivalence and Shiftable Functions]
  \label{corollary-shiftable:decidable}
Let $\alpha$ be a shiftable aggregation function over $\I^k$. 
Then for disjunctive $\alpha$-queries with comparisons ranging over $\I$ 
the bounded equivalence problem is decidable.
\end{corollary}

\begin{corollary}[Decidable Query Classes]
        \label{corollary-decidable:query:classes}
For the classes of disjunctive 
$\MAX$, $\SUM$, $\PROD$, $\AVG$, 
$\CNTD$, $\COUNT$, $\PRTY$ and $\TOPTWO$ queries, 
bounded equivalence is decidable provided that the comparisons range
over $\rat$ or $\intg$.
\end{corollary}

\begin{proof}
For the classes of disjunctive $\MAX$,  
$\CNTD$, $\COUNT$, $\PRTY$ and $\TOPTWO$, the claim follows 
directly from 
Corollary~\ref{corollary-shiftable:decidable}. For 
disjunctive $\SUM$ and $\AVG$ queries, decidability follows from
Proposition~\ref{prop:sum:avg:order-decidable} and
Theorem~\ref{t:bounded:equivalence}. Similarly, for disjunctive
$\PROD$ queries, 
the claim follows from Proposition~\ref{prop:prod:order-decidable} and
Theorem~\ref{t:bounded:equivalence}.
\end{proof}

In Theorem~\ref{t:bounded:equivalence} we reduced order-decidability
to bounded equivalence. 
A point of interest is that in our reduction we only used positive
queries. 
Therefore, negation in  queries $q$ and $q'$ does not affect 
decidability of bounded equivalence of $q$ and $q'$.

\begin{corollary}[Bounded Equivalence of Queries Without Negation]
The bounded equivalence
problem is decidable for {\em positive\/}
disjunctive $\alpha$-queries with comparisons ranging
over $\I$ if and only if the bounded equivalence
problem is decidable for 
disjunctive $\alpha$-queries {\em with negation\/} and comparisons ranging
over $\I$.
\end{corollary}


\section{Decomposition Principles}
        \label{sec-decomposition}

Levy and Sagiv~\cite{Levy:Sagiv-Queries:Independent:Updates-VLDB}
have shown that two disjunctive non-aggregate queries are equivalent if
they are equivalent over all databases whose carrier is not greater than
the size of the queries.
For non-aggregate queries this is not surprising since an answer by a
query~$q$ depends only on a single assignment satisfying~$q$.
Hence, if over some database $\D$, 
by means of the assignment~$\gamma$,
the query~$q$ returns the tuple $\tpl d$, 
but $q'$ does not return $\tpl d$,
then we can construct a database $\D_0 \incl \D$ that contains only 
constants occurring in $q$, $q'$ and $\gamma$ such that 
$\tpl d \in \DB q{\D_0}$ and $\tpl d \notin \DB{q'}{\D_0}$.

For aggregate queries this argument cannot be applied since the
results of a query are the amalgamation of many single results that may
involve arbitrarily many constants in the database.
Nevertheless, for queries with an idempotent monoid or a group aggregation
function we can reduce equivalence over arbitrary databases to
equivalence over small databases.

As a first step, we formulate {\em decomposition principles\/} for
these two classes of functions.
Such a principle provides a method to compute the value of an
aggregation over a union of sets of assignments from aggregations over
the sets themselves and possibly some of their subsets.

Note that the \quotes{$\sum$} in the equation below
is the extension of the
idempotent monoid operation.  
In the case of $\MAX$, for instance, the right hand side of the
equation becomes  $\max_{i=1}^k \, (\MAX(y)\downarrow{A_i})$.

\begin{proposition}[Idempotent Decomposition Principle]
        \label{prop-idempotent:decomposition:principle}
Let $\alpha$ be an idempotent monoid aggregation function and
$(A_i)_{i=1}^k$ a family of sets of assignments, 
all defined for $\tpl y$.  
Then
\begin{eqnarray}
        \label{eqn-idempotent:decomposition}
\mbox{$
        \alpha(\tpl y)\downarrow \bigcup_{i=1}^k A_i$}
     = 
\mbox{$
        \sum_{i=1}^k \, (\alpha(\tpl y)\downarrow A_i)$.}
\end{eqnarray}
\end{proposition}

\begin{proof} 
Let $(A_i)_{i=1}^k$ be a family of sets of assignments.
On the left hand side of Equation~\eqref{eqn-idempotent:decomposition},
first the union of the $A_i$ is taken and $\alpha$ is applied afterwards.
On the right hand side, $\alpha$ is applied 
first to (a projection of) each set $A_i$, and then the sum of the results
is taken.
Of course, it is possible for two different sets $A_i$ and $A_j$ 
to contain common elements. 
Note that on the left hand side, these duplicates are removed, 
whereas in the right side they are preserved. 

The aggregation function $\alpha$ is associative and commutative.
Therefore, the order in which $\alpha$ is applied to the assignments
is not important.
Hence, the only difference between the two sides of 
Equation~\eqref{eqn-idempotent:decomposition} that might 
affect the result is that 
on the right side the same assignment may appear in the 
summation several times. 

However, because of the associativity and commutativity of the monoid
operation, we may first sum such assignments with themselves.
Since $\alpha$ is idempotent, the final result is the same
as if each element had occurred only once.
Therefore, Equation~\eqref{eqn-idempotent:decomposition} holds.
\end{proof}

Before we treat the case of group aggregation functions, we remind the
reader of the 
well-known Principle of Inclusion and Exclusion for computing the
cardinality of a union of sets.
It says that for any finite family of sets 
$(A_i)_{i=1}^k$ we have
\begin{align}
        \label{eqn-inclusion:exclusion}
\textstyle{\bigcard{\bigcup_{i=1}^k A_i}}
        &= 
\textstyle{\sum_{i=1}^k \card{A_i} 
             - \sum_{i<j} \card{A_i \cap A_j} + \cdots +} \myvpace 
            \textstyle{(-1)^{k-1} \card{\bigcap_{i=1}^k A_i}}. \myvpace
\end{align} 

For group aggregation functions, we can generalize
Equation~\eqref{eqn-inclusion:exclusion}. 
In the following decomposition principle, the \quotes{$-$}-sign
denotes the inverse with respect to the group operation.
Note that for $\alpha = \COUNT$, Equation~\eqref{eqn-group:decomposition}
simplifies to Equation~\eqref{eqn-inclusion:exclusion},
since $(\COUNT \downarrow A) = \card A$ for every set of assignments $A$.

\begin{proposition}[Group Decomposition Principle]
                \label{prop-group:decomposition:principle}
Suppose that $\alpha$ is a group aggregation function and
$(A_i)_{i=1}^k$ is a finite family of sets of assignments, 
all defined for $\tpl y$.  Then
\eat{\begin{eqnarray}
 \renewcommand{\arraystretch}{5}
        \label{eqn-group:decomposition}
\mbox{$
\alpha(\tpl y) \downarrow \bigcup_{i=1}^k A_i$}
 & = &
        \mbox{%
            \renewcommand{\baselinestretch}{1.5}
            \small\normalsize
            $\sum_{i=1}^k\, (\alpha(\tpl y) \downarrow A_i)\,  
                - \,$} \\ 
&& \mbox{%
\renewcommand{\baselinestretch}{1.5}
            \small\normalsize
$\sum_{i<j}\, (\alpha(\tpl y) 
                        \downarrow {A_{i} \cap A_{j}}) \,+ \cdots +$} 
        \nonumber \\
&& \mbox{%
            \renewcommand{\baselinestretch}{1.5}
            \small\normalsize
            $(-1)^{k-1}\, (\alpha(\tpl y) 
                        \downarrow \bigcap_{i=1}^k A_i)$.}\nonumber
\label{equation:inclusion:exclusion}
\end{eqnarray}
}

\begin{align}
\begin{split}
        \label{eqn-group:decomposition}
\textstyle{\alpha(\tpl y) \downarrow \bigcup_{i=1}^k A_i}
  = {}& 
\textstyle{\sum_{i=1}^k\, (\alpha(\tpl y) \downarrow A_i)\,  
                - \,} 
\textstyle{\sum_{i<j}\, (\alpha(\tpl y) 
                        \downarrow {A_{i} \cap A_{j}}) \,+ \cdots +}
                                \myvpace\\ 
&
  \textstyle{(-1)^{k-1}\, (\alpha(\tpl y) 
                        \downarrow \bigcap_{i=1}^k A_i).} 
\end{split}
\end{align}
\end{proposition}

\begin{proof}
Equation~\eqref{eqn-group:decomposition} can be proved 
in the same fashion that the Principle of Inclusion and Exclusion 
is proved. 
Note that the right hand side of the equation is well defined 
since $\alpha$ takes values in an abelian group.

Let $\gamma$ be an assignment. 
If $\gamma$ is not in $A_i$ for any $i$, 
then clearly, $\gamma$ does not affect the value 
on the left or the right hand side 
of Equation~\eqref{eqn-group:decomposition}.

Suppose that $\gamma$ is in $r$ different sets $A_i$. 
The assignment $\gamma$ contributes $\gamma(\tpl y)$ once 
to the union of sets to which $\alpha$ is applied 
on the left hand side on the Equation. 
On the right hand side of the Equation, 
since $\gamma$ is in $r$ different sets $A_i$,
the value $\gamma(\tpl y)$ 
is added and subtracted (possibly) several times. 
To prove equality of the two sides, 
it is sufficient to show that on the right hand side
$\gamma(\tpl y)$ is added exactly one more time than it is subtracted. 

Since $\gamma$ occurs in $r$ sets $A_i$, 
it occurs in $\binom r 2$ intersections of two sets,
$\binom r 3$ intersections of three sets, etc.
Thus, on the right hand side,
$\gamma(\tpl y)$ is added $r$ times, 
then subtracted $\binom r 2$ times, 
then added $\binom r 3$ times, etc., 
In total, $\gamma$ contributes the tuple $\gamma(\tpl y)$ 
\[ 
  \binom r 1 - \binom r 2 + \cdots + (-1)^{r-1} \binom r r  = 1
\]
times to the right hand side of 
Equation~\eqref{eqn-group:decomposition}, 
as required.
\end{proof}

Because of the above two propositions,
we say that idempotent monoid and group aggregation functions are 
{\em decomposable}.

\section{Reducing Equivalence to Local Equivalence}
        \label{sec-reduction}

We now show that for queries with decomposable aggregation functions,
local equivalence implies equivalence. 
To this end we first show  that, given two queries and a database, we
can identify small subsets of the database, such that the satisfying
assignments over the database are the union of the satisfying
assignments over the subsets.
Then we apply the decomposition principles to conclude that the queries
return the same result over the original database
from the fact that the queries return the same results over the
small databases.

Let $q_1$ and $q_2$ be disjunctive queries,
$\D$ be a database,
and $\tpl d$ be a tuple of constants.
Let $(\D_i)_{i = 1}^k$ be a family of databases with $\D_i \incl \D$
for all $i = 1,\ldots,k$.
Then $(\D_i)_{i = 1}^k$ is a {\em decomposition\/} of $\D$ 
with respect to $q_1$, $q_2$ and $\tpl d$ if the following holds:
\begin{enumerate}
\item $\card{\carr{\D_i}} \leq \tsize{q_1,q_2}$ 
                              \ \ for all $i = 1,\ldots,k$; 
                                                    \label{prop:decomp:1}
\item $\ASSd{q_j}{\D}  =  \bigcup_{i=1}^k \ASSd{q_j}{\D_i}$
                              \ \ for $j = 1$,~$2$; 
                                                    \label{prop:decomp:2}
\item $\bigcap_h \ASSd{q_j}{\D_{i_h}} = 
                        \ASSd{q_j}{\bigcap_h \D_{i_h}}$
                              \ \ for $j = 1$,~$2$ and 
                              for all subfamilies $(\D_{i_h})_h$ of
                              $(\D_i)_i$.
                                                    \label{prop:decomp:3}
\end{enumerate}
The first condition means that, intuitively, 
the data\-bases $\D_i$ are small.
The second condition says that for each $q_j$, $j = 1$,~$2$,
we obtain exactly the satisfying assignments over $\D$ 
that return $\tpl d$ 
if we evaluate $q_j$ over each $\D_i$ separately 
and select the assignments that return $\tpl d$ over $\D_i$.
The third condition says that for each $q_j$,
in order to obtain the intersection of
the assignment sets $\ASSd{q_j}{\D_{i_h}}$, 
it suffices to evaluate $q_j$ over the intersection 
of the databases $\D_{i_h}$.


We will prove that given a pair of queries $q$ and $q'$, a database $\D$
and a tuple $\tpl d$, there exists a decomposition of $\D$ with
respect to $q$, $q'$ and $\tpl d$. To this end, we will first prove a
series of lemmas. 

We consider queries $q$ and $q'$ defined as
\begin{align*}
        q(\tpl x, \alpha(\tpl y)) &\qif \bigvee_{i\in I} P_i \AND N_i \AND C_i\\
        q'(\tpl x, \alpha(\tpl y')) &\qif \bigvee_{j\in J} P'_j \AND N'_j 
                                           \AND C'_j
\end{align*}
where $P_i$ and $P'_j$ are conjunctions of positive relational atoms,
$N_i$ and $N'_j$ are conjunctions of negated relational atoms and
$C_i$ and $C'_j$
are conjunctions of comparisons. We use $A_i$ as a shorthand for
$P_i\AND N_i \AND C_i$ and we use $A'_j$ as a shorthand for $P'_j\AND
N'_j\AND C'_j$. 

Let $\D$ be a database and let $\tpl d$ be a tuple. We must show that
there exists a decomposition of $\D$ with respect to $q$, $q'$ and $\tpl d$. 

We create a decomposition of $\D$ with respect to $q$,~$q'$ and $\tpl
d$ in a two-step process. We first
create databases out of the satisfying assignments of $q$ and of $q'$ into
$\D$ that retrieve $\tpl d$. Next, we extend these databases using the
procedure {\sc Extend\_Database\/} to prevent them from satisfying
negated atoms that were not satisfied in $\D$.

\begin{figure}[!t]
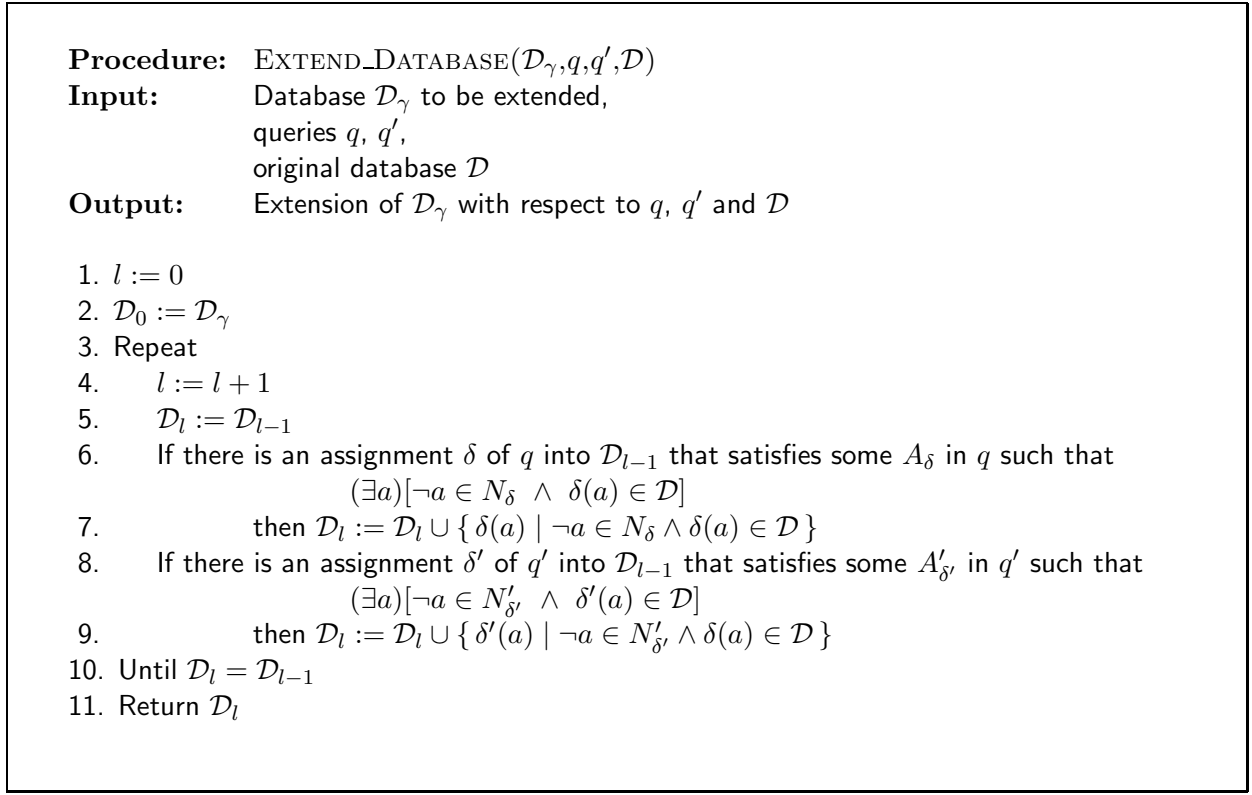

{\sf
\boxfigureone{%
\begin{tabbing} 
AA12. \= AAAAA\= AAAAA\= AAAAA\= AAAAA\= AAAAA\= AAAAA\= \kill
{\bf Procedure:} \>\> {\sc Extend\_Database($\D_\gamma$,$q$,$q'$,$\D$)} \\
{\bf Input:}     \>\> Database $\D_\gamma$ to be extended, \\
                 \>\> queries $q$, $q'$, \\
                 \>\> original database $\D$\\
{\bf Output:}    \>\> Extension of $\D_\gamma$ with respect to $q$,
                      $q'$ and $\D$\\ 
\\
\ 1. $l := 0$ \\
\ 2. $\D_0 := \D_\gamma$ \\
\ 3. Repeat \\
\ 4. \> $l := l + 1$ \\
\ 5. \> $\D_l := \D_{l-1}$ \\
\ 6. \> {If} there is an assignment $\delta$ of $q$ into
        $\D_{l-1}$ that satisfies some $A_{\delta}$ in $q$ such that  \\
     \>\>\>
        $(\exists a)[\neg a\in N_{\delta}\ \wedge\ \delta(a)\in \D]$ \\
\ 7. \>\> {then} $\D_l := \D_{l}\cup \set{\delta(a) \mid \neg a \in
  N_{\delta}\wedge \delta(a)\in\D}$  \\
\ 8. \>{If} there is an assignment $\delta'$ of $q'$ into
$\D_{l-1}$ that  satisfies some $A'_{\delta'}$ in $q'$ such that \\
\>    \> \> 
        $(\exists a)[\neg a\in N'_{\delta'}\ \wedge\ \delta'(a)\in \D]$ \\
\ 9. \>\>{then} $\D_l := \D_l\cup \set{\delta'(a)  \mid  \neg a \in N'_{\delta'}\wedge \delta(a)\in\D}$ \\
10. {Until} $\D_l = \D_{l-1}$ \\
11. Return $\D_l$ 
\end{tabbing}}}\caption{Procedure used to extend a database.}\label{fig:extend:database}
\end{figure}

Recall that $\ASSd{q}{\D}$ is the set of satisfying assignments from
$q$ into $\D$ that retrieve $\tpl d$.
We denote the disjunct of $q$ satisfied by $\gamma\in \ASSd{q}{\D}$ as $A_\gamma$.
For each $\gamma\in \ASSd{q}{\D}$, we define a database
        \[\D_{\gamma} := \set{\gamma(a) \mid a\in P_{\gamma}}.\] 
We use this notation since we consider a database  to be a set of ground 
positive relational  atoms. 
Note that $\D_\gamma$ satisfies the positive atoms in $A_{\gamma}$
with respect to the 
assignment $\gamma$. However, we must extend the databases $\D_\gamma$
to ensure that $\D_\gamma$ does not satisfy negated atoms that were
not satisfied in $\D$.  We now create a database $\D_\gamma^*$ out of 
$\D_\gamma$ using the procedure {\sc Extend\_Database\/} presented in 
Figure~\ref{fig:extend:database}.\footnote{This process does not
 necessarily uniquely determine the database $D_\gamma^*$. However, this is 
not important for our proof. }
 Formally, we define 
$\D_\gamma^*:= \mbox{\sc Extend\_Database}(\D_\gamma,q,q',\D)$.

In a similar fashion, we create databases $\D_{\gamma'}$ out of the 
satisfying assignments $\gamma'\in \ASSd{q'}{\D}$ of $q'$ into $\D$ 
that retrieve $\tpl d$. 
As above, these databases are extended to derive databases $\D_{\gamma'}^*$
using the procedure {\sc Extend\_Database}. 

We now define 
\begin{equation}
        \label{eqn-decomposition:delta}
  \Delta := \set{\D_\gamma^* \mid  \gamma\in \ASSd{q}{\D}} \cup 
            \set{\D_{\gamma'}^* \mid \gamma'\in\ASSd{q'}{\D}}.
\end{equation}

We present a series of lemmas that will enable us to prove that
$\Delta$ is a decomposition of $\D$ 
w.r.t.\ $q$, $q'$ and $\tpl d$.  
We first note that clearly for all $\D^* \in \Delta$, 
it holds that $\D^*\subseteq \D$.

The first lemma states that 
the databases in $\Delta$ have the correct number of constants, i.e., 
that Property~\ref{prop:decomp:1} of decompositions 
holds for $\Delta$.

\begin{lemma}[Size of Databases]
       \label{lemma:size:databases}
For all databases $\D^*\in\Delta$, it holds that 
$\card{\carr{\D^*}} \leq \tsize{q,q'}$.
\end{lemma}

\begin{proof}
Consider a database $\D_\gamma^*\in\Delta$. 
Clearly, $\D_\gamma$ contains at most $\tsize{q}$ constants. 
Note that when an atom is added during the procedure, 
the constants appearing in the atom 
must have already appeared in the database $\D_\gamma$, 
or must appear in $q$ or $q'$. 
This follows since
the queries are safe and 
all variables in negated atoms must also appear in positive atoms (or
be equated to variables appearing in positive atoms). 
Thus, $\D_\gamma^*$ contains at most $\tsize{q,q'}$ constants.

Similarly, 
one can show that for $\D_{\gamma'}^*\in\Delta$, 
it holds that $\card{\carr{\D_{\gamma'}^*}} \leq \tsize{q,q'}$. 
Thus, it easily follows that for all $\D^*\in\Delta$,  
we have $\card{\carr{\D^*}} \leq \tsize{q,q'}$.
\end{proof}

We show Property~\ref{prop:decomp:2} of decompositions for $\Delta$.

\begin{lemma}[Assignments into $\D$ and $\Delta$]
       \label{lemma:assignments}\mbox{} 
The following relationships hold between 
the assignments of $q$ and $q'$ into $\D$ 
and into databases in $\Delta$:
\begin{enumerate}
\item  $\ASSd{q}{\D}  =  \bigcup_{\D^*\in\Delta} \ASSd{q}{\D^*}$; 
                \label{lemma:item:q}
\item  $\ASSd{q'}{\D}  =  \bigcup_{\D^*\in\Delta} \ASSd{q'}{\D^*}$.
                \label{lemma:item:q-tag}
\end{enumerate}
\end{lemma}

\begin{proof}
We only prove Part~\ref{lemma:item:q}. 
Part~\ref{lemma:item:q-tag} can be shown analogously.
We show the set equality in Part~\ref{lemma:item:q}
by proving two inclusions. 

\quotes{$\incl$}\ \ 
Suppose that $\gamma\in\ASSd{q}{\D}$.
It is enough to show that $\gamma$ satisfies $q$ over $\D_\gamma^*$,
which entails that $\gamma\in \ASSd{q}{\D_\gamma^*}$.

Let $a$ be a positive relational atom 
in the conjunct $A_{\gamma}$.
Then 
 $\gamma(a)\in \D_\gamma$ by definition. 
Clearly, $\D_\gamma\subseteq\D_\gamma^*$, and thus, 
 $\gamma(a)\in\D_\gamma^*$. 
If $\neg b$ is a negated relational atom in  $A_{\gamma}$
then $\gamma(b) \not\in \D$. 
Otherwise, $\gamma$ would not be 
a satisfying assignment of $q$ in $\D$. 
According to the definition of $\D_\gamma^*$, it holds that
 $\D_\gamma^*\subseteq \D$, and 
therefore, $\gamma(b)\not\in \D_\gamma^*$. 
The satisfaction of $C_{\gamma}$ 
 (the comparisons in $A_{\gamma}$) 
depends only on $\gamma$ and not on any database. 
Thus, $\gamma$ is a satisfying assignment of $q$ over $\D_\gamma^*$. 

\quotes{$\supseteq$}\ \ 
It suffices to show that for all $\D^*\in\Delta$ it holds that 
 $\ASSd{q}{\D}  \supseteq  \ASSd{q}{\D^*}$.
Suppose that 
 $\gamma\in\ASSd{q}{\D^*}$. 
We show that $\gamma$ is a satisfying assignment of $q$ over $\D$. 

Suppose that $\gamma$ satisfies the conjunct 
$A_{\gamma}$ of $q$ in $\D^*$. 
Consider a literal $l$ in $A_{\gamma}$. 
If $l$ is a positive relational atom then $\gamma(l)\in \D^*$. 
We know that $\D^*\subseteq \D$, thus, $\gamma(l)\in \D$. 
Suppose that $l$ is a negated relational atom of the form $\neg b$, 
and suppose, by way of contradiction, that $\gamma(b)\in\D$. 
Then $\gamma$ satisfies the condition in line 6 of the procedure 
{\sc Extend\_Database\/} presented above. 
Thus, $\gamma(b)$ would have been added to $\D^*$ 
in contradiction to the fact that 
$\gamma$ is a satisfying assignment of $q$ into $\D^*$. 
Finally, note that the satisfaction of comparisons depends only
on the assignment, and not on the database.
Thus, $\gamma$ is a satisfying assignment of $q$ over $\D$. 
\end{proof}

In Lemma~\ref{lemma:assignments:intersections},
Property~\ref{prop:decomp:3} of decompositions is proved for $\Delta$.

\begin{lemma}[Assignments into Intersections]
        \label{lemma:assignments:intersections}
The following relationships hold between intersections of sets of assignments
and intersections of sets of databases:
\begin{enumerate}
\item $\bigcap_h \ASSd{q}{\D^*_h} = 
                        \ASSd{q}{\bigcap_h \D^*_h}$;
        \label{lemma:item:intersection:q}
\item $\bigcap_h \ASSd{q'}{\D^*_h} = 
                        \ASSd{q'}{\bigcap_h \D^*_h}$
        \label{lemma:item:intersection:q-tag}
\end{enumerate}
for all subfamilies $\D^*_h$ of $\Delta$.
\end{lemma}

\begin{proof}
We only prove Part~\ref{lemma:item:intersection:q}. 
Part~\ref{lemma:item:intersection:q-tag} can be shown analogously.
We show the set equality in Part~\ref{lemma:item:intersection:q}
by proving two inclusions. 

\quotes{$\supseteq$}\ \ 
Let $\gamma$ be an assignment in  $\ASSd{q}{\bigcap \D^*_h}$. 
Suppose that $\gamma$ satisfies 
$A_{\gamma}$ of $q$ in $\bigcap \D^*_h$.
Satisfaction of $C_{\gamma}$ is dependent only on $\gamma$.
Let $a$ be a positive atom in $A_{\gamma}$. 
The atom $\gamma(a)$ appears in $\bigcap \D^*_h$
and thus, $\gamma(a)$ appears in $\D^*_h$ for all $h$. 
Thus, $\gamma$ satisfies 
the positive atoms of $A_{\gamma}$ in each of the $\D^*_h$. 
Now, let $l$ be a negated atom in $A_{\gamma}$ of the form $\neg b$.
Clearly, $\gamma(b)\not\in\bigcap \D^*_h$. 
Suppose, by way of contradiction, that $\gamma(b)\in\D^*_h$ for some $h$. 
Then $\gamma(b)\in \D$, since $\D^*_h\subseteq \D$. 
However, it follows that 
we would have added $\gamma(b)$ to $\D^*_h$ for all $h$, 
since $\gamma$ satisfies the condition in line 6 of {\sc Extend\_Database}. 
Thus, $\gamma(b)\in \bigcap \D^*_h$ in contradiction to the assumption. 
This proves that $\gamma \in \bigcap \ASSd{q}{\D^*_h}$.
Hence, we have shown that
every $\gamma$ that is an element of the right hand side of
Equation~\eqref{lemma:item:intersection:q} 
is also an element of the left hand side.


\quotes{$\subseteq$}\ \ 
Suppose that $\gamma\in\bigcap \ASSd{q}{\D^*_h}$.
Then $\gamma\in \ASSd{q}{\D^*_h}$ for all $h$. 
Let $A_{\gamma}$ be a conjunct such that 
$\gamma$ satisfies $A_{\gamma}$ in $\D^*_h$ for all $h$. 
(Recall the definition of $\ASSqD$ in Section~\ref{subsec-sem:agg:query},
 where we have assumed that each $\gamma$ carries a label, recording
 which disjunct of $q$ it satisfies.)
Once again, satisfaction of $C_{\gamma}$ is dependent only on $\gamma$.
Consider a positive relational atom $a$ in $A_{\gamma}$. 
Then $\gamma(a)\in \D^*_h$ for all $h$. 
Thus, $\gamma(a)\in \bigcap \D^*_h$. 
Similarly, consider a negated atom $l$ of the form $\neg b$ in $A_{\gamma}$.
Then $\gamma(b)\not\in \D^*_h$ for all $h$, and thus, 
 $\gamma(b)\not\in \bigcap \D^*_h$. 
This proves that $\gamma\in \ASSd{q}{\bigcap \D^*_h}$. 
Hence, the second inclusion holds as well.
\end{proof}

We can now prove our theorem about the existence of decompositions.

\begin{theorem}[Existence of Database Decompositions]
  \label{T:decomposition}
Let $q$ and $q'$ be a pair of disjunctive queries,
let $\D$ be a database, and let
$\tpl d$ be a tuple of constants from $\D$.
Then there exists a decomposition of $\D$ \wrt $q$,~$q'$ and $\tpl d$.
\end{theorem}

\begin{proof}
From Lemmas~\ref{lemma:size:databases},~\ref{lemma:assignments} 
and~\ref{lemma:assignments:intersections} it follows that $\Delta$ 
as defined in Equation~\eqref{eqn-decomposition:delta} is
a decomposition of $\D$ \wrt $q$, $q'$ and $\tpl d$ as required.
\end{proof}

\eat{
Finally, we give the proof that local equivalence implies equivalence
for aggregate queries with idempotent monoid functions.

\begin{theorem}[Reduction to Local Equivalence]
Let $\alpha$ be an idempotent monoid aggregation function, 
and let $q$ and $q'$ be disjunctive $\alpha$-queries. 
Then $q$ and $q'$ are equivalent if and only if they are 
locally equivalent.
\end{theorem}

\begin{proof}
We only have to show that local equivalence implies equivalence.
Suppose therefore that $q$ and $q'$ agree on all databases whose
carrier has at most $\tsize{q,q'}$ elements.
Let $\D$ be any database and $\tpl d$ be a tuple of constants.
It suffices to show that 
\begin{eqnarray*}
\alpha(\tpl y) \downarrow \ASSd{q}{\D} = 
        \alpha(\tpl y) \downarrow \ASSd{q'}{\D}.
\end{eqnarray*}

Let $(\D_i)_{i=1}^k$ be a {\em decomposition\/} of $\D$ with respect
to $q$, $q'$ and to $\tpl d$.
If $\alpha$ is an idempotent monoid function, we apply 
Proposition~\ref{prop-idempotent:decomposition:principle}, which yields
\begin{subequations}
\begin{align}
\alpha(\tpl y) \downarrow \ASSd{q}{\D}  
        &=  
        \textstyle{\alpha(\tpl y) \downarrow \bigcup_{i=1}^k \ASSd{q}{\D_i}}
                \myvpace
                \label{eqn-decomposition:union:q} \\
        &=  
        \textstyle{\sum_{i=1}^k \, (\alpha(\tpl y) \downarrow \ASSd{q}{\D_i})}
                \myvpace
                \label{eqn-idempotent:principle:q} \\
        &=  
        \textstyle{\sum_{i=1}^k \, (\alpha(\tpl y) \downarrow \ASSd{q'}{\D_i})}
                \myvpace
                \label{eqn-local:equivalence:idempotent} \\
        &=  
        \textstyle{\alpha(\tpl y) \downarrow \bigcup_{i=1}^k \ASSd{q'}{\D_i}}
                \myvpace
                \label{eqn-idempotent:principle:q:prime} \\
        &= 
        \textstyle{\alpha(\tpl y) \downarrow \ASSd{q'}{\D}} \myvpace,
                \label{eqn-decomposition:union:q:prime}
\end{align}
\end{subequations}
where Equations~(\ref{eqn-decomposition:union:q}) 
        and~(\ref{eqn-decomposition:union:q:prime}) hold 
because of Property~2 of decompositions, 
Equations~(\ref{eqn-idempotent:principle:q}) 
        and~(\ref{eqn-idempotent:principle:q:prime}) hold 
because of Proposition~\ref{prop-idempotent:decomposition:principle}, and 
Equation~(\ref{eqn-local:equivalence:idempotent}) holds because
$q$ and $q'$ are locally equivalent and the databases $\D_i$ contain
at most $\tsize{q,q'}$ constants.

\end{proof}
}


Finally, we reduce equivalence to local equivalence. 

\begin{theorem}[Reduction to Local Equivalence]
    \label{T:reduction}
Let $\alpha$ be a decomposable aggregation function, 
and let $q$ and $q'$ be disjunctive $\alpha$-queries. 
Then $q$ and $q'$ are equivalent if and only if they are 
locally equivalent.
\end{theorem}

\begin{proof}
We only have to show that local equivalence implies equivalence.
Suppose therefore that $q$ and $q'$ agree on all databases whose
carrier has at most $\tsize{q,q'}$ elements.
Let $\D$ be any database and $\tpl d$ be a tuple of constants.
It suffices to show that 
\begin{equation*}
\alpha(\tpl y) \downarrow \ASSd{q}{\D} = 
        \alpha(\tpl y) \downarrow \ASSd{q'}{\D}.
\end{equation*}

Let $(\D_i)_{i=1}^k$ be a decomposition of $\D$ 
\wrt $q$, $q'$ and to $\tpl d$.
If $\alpha$ is an idempotent monoid function, we apply 
Proposition~\ref{prop-idempotent:decomposition:principle}, which yields
\begin{subequations}
\begin{align}
\alpha(\tpl y) \downarrow \ASSd{q}{\D}  
        &=  
        \alpha(\tpl y) \downarrow \textstyle{\bigcup_{i=1}^k} \ASSd{q}{\D_i} 
                \label{eqn-decomposition:union:q} \myvpace \\
        &=  
        \textstyle{\sum_{i=1}^k} \, (\alpha(\tpl y) \downarrow \ASSd{q}{\D_i}) 
                \label{eqn-idempotent:principle:q} \myvpace\\
        &=  
        \textstyle{\sum_{i=1}^k} \, (\alpha(\tpl y) \downarrow \ASSd{q'}{\D_i})
                \label{eqn-local:equivalence:idempotent} \myvpace\\
        &= 
        \alpha(\tpl y) \downarrow \textstyle{\bigcup_{i=1}^k} \ASSd{q'}{\D_i} 
                \label{eqn-idempotent:principle:q:prime} \myvpace \\
        &=
        \alpha(\tpl y) \downarrow \ASSd{q'}{\D},\myvpace
                \label{eqn-decomposition:union:q:prime}
\end{align}
\end{subequations}
where Equations~\eqref{eqn-decomposition:union:q} 
        and~\eqref{eqn-decomposition:union:q:prime} hold 
because of Property~2 of decompositions, 
Equations~\eqref{eqn-idempotent:principle:q}
        and~\eqref{eqn-idempotent:principle:q:prime} hold 
because of Proposition~\ref{prop-idempotent:decomposition:principle}, and 
Equation~\eqref{eqn-local:equivalence:idempotent} holds because
$q$ and $q'$ are locally equivalent and the databases $\D_i$ contain
at most $\tsize{q,q'}$ constants.

If $\alpha$ is a group aggregation function, we apply 
Proposition~\ref{prop-group:decomposition:principle}, 
which yields the equations 
\begin{subequations}
\begin{align}
\alpha(\tpl y) \downarrow \ASSd{q}{\D}  
        &= 
        \textstyle{\alpha(\tpl y) \downarrow \bigcup_{i=1}^k \ASSd{q}{\D_i}}
                \myvpace 
                \label{eqn-group:decomposition:union:q}\\
        &=  
        \textstyle{\sum_{i=1}^k \, (\alpha(\tpl y) \downarrow \ASSd{q}{\D_i}) 
        - \cdots + 
        (-1)^{k-1}\, (\alpha(\tpl y) \downarrow 
                        \bigcap_{i=1}^k \ASSd{q}{\D_i})} \myvpace
                \label{eqn-group:principle:q}\\
        &= 
        \textstyle{\sum_{i=1}^k \, (\alpha(\tpl y) \downarrow \ASSd{q}{\D_i}) 
        - \cdots + 
        (-1)^{k-1}\, (\alpha(\tpl y) \downarrow 
                         \ASSd{q}{\mbox{$\bigcap_{i=1}^k\D_i$}})}
        \myvpace
                \label{eqn-group:decomposition:inters:q}\\
        &=  
        \textstyle{\sum_{i=1}^k \, (\alpha(\tpl y) \downarrow \ASSd{q'}{\D_i}) 
        - \cdots + 
        (-1)^{k-1}\, (\alpha(\tpl y) \downarrow 
                         \ASSd{q'}{\mbox{$\bigcap_{i=1}^k\D_i$}})}
        \myvpace 
                \label{eqn-local:equivalence:group}\\
        &= 
        \textstyle{\sum_{i=1}^k \, (\alpha(\tpl y) \downarrow \ASSd{q'}{\D_i}) 
        - \cdots + 
        (-1)^{k-1}\, (\alpha(\tpl y) \downarrow 
                        \bigcap_{i=1}^k \ASSd{q'}{\D_i})} \myvpace
                      \label{eqn-group:decomposition:inters:q:prime}\\
        &= 
        \textstyle{\alpha(\tpl y) \downarrow \bigcup_{i=1}^k \ASSd{q'}{\D_i}}
                \myvpace 
                \label{eqn-group:principle:q:prime}\\
        &= 
        \textstyle{\alpha(\tpl y) \downarrow \ASSd{q'}{\D}}\myvpace 
                \label{eqn-group:decomposition:union:q:prime}
\end{align}
\end{subequations}
where Equations~\eqref{eqn-group:decomposition:union:q}
        and~\eqref{eqn-group:decomposition:union:q:prime} hold 
because of Property~2 of decompositions, 
Equations~\eqref{eqn-group:principle:q}
        and~\eqref{eqn-group:principle:q:prime} hold 
because of Proposition~\ref{prop-group:decomposition:principle}, 
Equations~\eqref{eqn-group:decomposition:inters:q} 
        and~\eqref{eqn-group:decomposition:inters:q:prime} hold 
because of Property~3 of decompositions, and 
Equation~\eqref{eqn-local:equivalence:group} holds because
$q$ and $q'$ are locally equivalent and the databases $\D_i$ contain
at most $\tsize{q,q'}$ constants.
\end{proof}

The aggregation function $\PROD$ is not a decomposable aggregation
function over $\rat$. However, $\PROD$ is decomposable over
$\ratnozero$, i.e.,  the rational numbers without the element 0. 
It turns out that this is sufficient
in order to reduce equivalence to local equivalence for $\PROD$, defined
over the rational numbers.

\begin{theorem}[Reduction to Local Equivalence for Product]
    \label{T:prod:reduction}
\ \ 
Suppose $q(\tpl x, \PROD(y))$ 
and \linebreak 
$q'(\tpl x, \PROD(y))$ are disjunctive $\PROD$-queries, 
defined over $\rat$. 
Then $q$ and $q'$ are equivalent if and only if they are 
locally equivalent.
\end{theorem}

\begin{proof}
As before, we only have to show that local equivalence implies
equivalence. Suppose therefore that $q$ and $q'$ agree on all
databases whose carrier has at most $\tsize{q,q'}$ elements.
Let $\D$ be any database and $\tpl d$ be a tuple of constants.
It suffices to show that 
\begin{equation}
        \label{eqn-equality:of:prod}
\PROD(y) \downarrow \ASSd{q}{\D} = 
        \PROD(y) \downarrow \ASSd{q'}{\D}.
\end{equation}

Let $(\D_i)_{i=1}^k$ be a decomposition of $\D$ 
\wrt $q$, $q'$ and to $\tpl d$. 
We distinguish between three cases.

\CASE{1}
  Suppose that there is an assignment $\gamma\in \ASSd q \D$ that 
  maps $y$ to 0. Then, $q$ retrieves the aggregate value 0 for $\tpl
  d$ over $\D$, i.e., $\PROD(y) \downarrow \ASSd{q}{\D} = 0$. By
  Property~2 of decompositions, there is a database 
  $\D_\gamma\in (\D_i)_{i=1}^k$ such that $\gamma\in \ASSd
  q{\D_\gamma}$. Note that $q$ returns the aggregate value 0 for $\tpl
  d$ over $\D_\gamma$. By Property~1 of decompositions, $\D_\gamma$ has at 
  most $\tsize{q,q'}$ elements. By our assumption, $q$ and $q'$ are
  locally equivalent. Therefore, $q'$ must return the aggregation
  value 0 for $\tpl d$ over $\D_\gamma$. Hence, by applying Property~2 of
  decompositions once more, we derive that $q'$ retrieves the
  aggregate value 0 for $\tpl d$ over $\D$. 

\CASE{2}
  Suppose that there is an assignment $\gamma\in \ASSd {q'} \D$ 
  that maps $y$ to 0. 
  By analogous arguments to the previous case, 
  we can show that both $q$ and $q'$ retrieve the
  aggregate value 0 for $\tpl d$ over~$\D$. 

\CASE{3}
  Suppose that there is no assignment in $\ASSd q \D$ that maps 
  $y$ to 0. Similarly, suppose that there is no assignment in $\ASSd
  {q'} \D$ that maps $y$ to 0. Then, the aggregation function 
  $\PROD$ could just as well have been defined over $\ratnozero$. 
  In this case, $\PROD$ is a decomposable aggregation function and the
  arguments used in Equations~\eqref{eqn-group:decomposition:union:q}
  through~\eqref{eqn-group:decomposition:union:q:prime} in the proof of
  Theorem~\ref{T:reduction} apply. Therefore, $q$ and $q'$ return the
  same aggregate value for $\tpl d$ over $\D$ as required.

Thus, we have proved that Equation~\eqref{eqn-equality:of:prod} holds
in all possible cases.
\end{proof}

The following result follows directly from Theorem~\ref{T:reduction}.

\begin{corollary}[Local Equivalence and Equivalence]
Suppose $\alpha$ is a decomposable aggregation function.  
If local equivalence is decidable for disjunctive $\alpha$-queries, 
then  equivalence is also decidable.   
\end{corollary}

In Section~\ref{sec-equivalence},
we have noted that bounded equivalence of $\alpha$-queries can be
checked in double exponential time if
ordered identities of the form defined in Formula~\eqref{e:decidable} can be
decided in double exponential time.
Hence, if $\alpha$ is also decomposable, we derive a double
exponential upper bound on checking for equivalence of
$\alpha$-queries.

From Theorems~\ref{T:reduction} and~\ref{T:prod:reduction} and
Corollary~\ref{corollary-decidable:query:classes} we derive the
following result.

\begin{corollary}[Decidable Query Classes]
Equivalence of disjunctive aggregate queries is decidable for the
aggregation functions
$\MAX$, $\TOPTWO$, $\COUNT$, $\PRTY$, and $\SUM$
over both the integers and the rational numbers. In addition,
equivalence of disjunctive $\PROD$-queries is decidable 
over the rational numbers. 
\end{corollary}


\newcommand{\naproj}[1]{\hat #1}
\newcommand{\QL}[1]{\Q\L(#1)}
\newcommand{\Lin}[1]{\L(#1)}

\section{Equivalence of Conjunctive Quasilinear Queries}
A positive conjunctive query $q$ is {\em linear\/} if no predicate occurs
more than once in $q$%
~\cite{Nutt:Et:Al-Equivalences:Among:Aggregate:Queries-PODS}. 
We generalize this by defining that a conjunctive query is {\em
quasilinear\/} if no predicate that occurs in a positive literal,
occurs more than once.
Thus, in a quasilinear query,
no predicate occurs in both a positive and a negated literal and
no predicate occurs more than once in a positive literal.
In this section we show that for a wide range of quasilinear queries,
equivalence is isomorphism.

In Section~\ref{sec:order-decid-aggr}.
we defined reduced sets of terms \wrt a complete ordering.
In a similar spirit, we now introduce 
reduced conjunctions of comparisons.
A conjunction of comparisons $C$ is {\em reduced\/} with respect to a
domain $\I$ if
\begin{itemize}
\item there are no variables $x$ and $y$ occurring in $C$ such that 
      $C \dommodels \I x = y$;
\item there is no variable $x$ occurring in $C$ such that 
      $C \dommodels \I x = d$  for a constant $d\in \I$.
\end{itemize}
We say that a conjunctive query is {\em reduced\/} with respect to
$\I$ if its comparisons are
reduced with respect to $\I$. If the domain is clear from the context,
we will simply say
that a query is reduced, without specifying the domain.

We have shown
in~\cite{Nutt:Et:Al-Equivalences:Among:Aggregate:Queries-PODS} 
that for any positive conjunctive query,
one can compute in polynomial time an equivalent reduced conjunctive query. 
This still holds when the query contains negated atoms.
Note that the head of the equivalent reduced query 
may contain constants,
even if the head of the original non-reduced query does not.

Let $q(\tpl s, \alpha(\tpl t)) \qif P \AND N\AND C$ and $q'(\tpl s',
\alpha(\tpl t')) \qif P' \AND N'\AND C'$ be  
conjunctive aggregate queries with comparisons, ranging over the
domain $\I$. We 
use $P$ and $P'$ to denote the positive atoms, $N$ and $N'$ to denote
the negated atoms, and $C$ and $C'$ to denote the comparisons.
A {\em \hom\/} from $q'$ to $q$ is a substitution $\theta$ of the
variables in $q'$ by terms in $q$ such that 
\begin{enumerate}
\item $\theta (\tpl s') = \tpl s$ and $\theta (\tpl t') = \tpl t$;
\item $\theta (a')$ is in $P$ for every positive relational atom $a'$ of $P'$;
\item $\theta (a')$ is in $N$ for every negated relational atom $a'$ of $N'$;
\item $C\dommodels \I \theta(s') \osps\rho \theta(t')$ for every 
      comparison $s'\osps\rho t'$ in $C'$.
\end{enumerate}
A \hom\ is an {\em isomorphism\/} if it is bijective and if its
inverse is also a \hom. 
The queries $q'$ and $q$ are {\em isomorphic\/} if there is an
isomorphism from $q'$ to $q$.
In~\cite{Nutt:Et:Al-Equivalences:Among:Aggregate:Queries-PODS} we have
also shown that reduced linear $\MAX$, $\COUNT$ and $\SUM$ queries
are equivalent if and only if they are isomorphic.
For queries with negated literals, we can generalize this result to
quasilinear queries.

We say that a class of queries $\Q$ is {\em proper\/}
if for satisfiable queries in $\Q$ equivalence implies isomorphism,
that is, 
if for any two satisfiable reduced queries $q$,~$q'\in\Q$ 
it is the case that $q$ and $q'$ are only equivalent 
if they are isomorphic. For every aggregation function we denote by
$\Lin\alpha$  the class of  linear $\alpha$-queries. 
Similarly, we denote by $\QL\alpha$ 
the class of  quasilinear $\alpha$-queries.

\begin{theorem}[Quasilinear and Linear Queries]
       \label{thm-negation:positive:isomorphism}
Let $\alpha$ be an aggregation function. Suppose that the class $\Lin
\alpha$ is proper. Then, the class $\QL \alpha$ is also proper.
\end{theorem}

\begin{proof}
Consider the satisfiable reduced queries
\begin{align*}
        q(\tpl s, \alpha(\tpl t)) &\qif P \AND N \AND C \\
        q'(\tpl s, \alpha(\tpl t)) &\qif P' \AND N' \AND C'
\end{align*}
where 
\begin{itemize}
        \item $P$ and $P'$ are conjunctions of positive relational atoms;
        \item $N$ and $N'$ are conjunctions of negated relational atoms;
        \item $C$ and $C'$ are conjunctions of comparisons.
\end{itemize}
Suppose that  $\Lin \alpha$ is proper.
Suppose that $q$ is not isomorphic to $q'$. 
We show that $q$ is not equivalent to $q'$. Note that we can assume
that $q$ and $q'$ have the same heads since otherwise they are
obviously not equivalent.

We introduce the {\em positive parts\/} of $q$ and $q'$ as the 
queries $\pos q$ and $\pos q'$, defined as
\begin{align*}
        \pos q(\tpl s, \alpha(\tpl t)) &\qif P \AND C\\
        \pos q'(\tpl s, \alpha(\tpl t)) &\qif P' \AND C' \, .
\end{align*}
We consider two cases.

\CASE{1}
Suppose that $\pos q$ is not isomorphic to $\pos{q'}$. 
Hence, $\pos q$ and $\pos{q'}$ are not equivalent.
Let $\D$ be a database for which 
$\pos q$ and $\pos{q'}$ return different values. 
We may assume, without loss of generality, that 
$\D$ only contains atoms with predicates appearing in $P$ or in $P'$. 

If there is a predicate $p$ that appears in an atom $P$, but not in $P'$, then
clearly $q$ and $q'$ cannot be equivalent, since we could create a
database that satisfies $q'$ and does not contain any atom with
predicate $p$. Similarly, if there is a predicate that appears in
$P'$, but not in $P$, then $q$ and $q'$ cannot be equivalent. Hence,
we may assume that the set of predicates of atoms in $P$ is identical
to the set of predicates of atoms in $P'$. 
Thus, there is no atom in $\D$ 
containing a predicate appearing in $N$ or $N'$. 

Thus, $\DBD{\pos q} = \DBD q$ and $\DBD{\pos{q'}}= \DBD{q'}$. 
We conclude that $\D$ is a counterexample 
for the equivalence of $q$ and $q'$. 

\CASE{2}
Suppose that $\pos q$ is isomorphic to $\pos{q'}$. 
Since $\pos q$ and $\pos{q'}$ are linear, 
there exists only one isomorphism between them, say $\theta$.
Note that $\theta$ is defined on all the variables in $q$, 
since $q$ is a safe query. 
By our assumption, $q$ is not isomorphic to $q'$. 
Thus, $\theta(N) \neq N'$. 
Suppose, without loss of generality, that $\neg a$ appears in $N$ and 
$\theta(\neg a)$ does not appear in $N'$. 
Let $\gamma$ be a mapping of the variables in $q$ to constants,
such that $\gamma$ is consistent with the comparisons in~$q$. 
We define a database 
$\D:=\set{\gamma(b) \mid b\in P} \cup \set{\gamma(a)}$. 
Clearly, $q$ does not return a grouping value 
for $\gamma(\tpl s)$ over $\D$
whereas $q'$ does return a grouping value for $\gamma(\tpl s)$. 
Thus, $q$ and $q'$ are not equivalent. 

This completes the proof.
\end{proof}


Now, it follows from our results
 in~\cite{Nutt:Et:Al-Equivalences:Among:Aggregate:Queries-PODS} 
that for quasilinear queries 
with the aggregate functions $\MAX$, $\SUM$ and $\COUNT$, 
equivalence boils down to isomorphism. 
In a similar fashion to the proofs there, we
can extend our results to additional aggregate functions. 

A bag $B$ is a {\em singleton\/} 
if it contains exactly one value. 
We say that an aggregation function $\alpha$ is a 
{\em singleton-determining\/} aggregation function, 
if for all singleton bags $B$ and $B'$ we have that 
\[ 
  \alpha(B) = \alpha(B') \iff B = B'. 
\] 
Clearly  $\MAX$, $\TOPTWO$,  $\SUM$, $\PROD$ and $\AVG$ 
are singleton-determining aggregation functions. Note that $\COUNT$
and $\PRTY$ are nullary aggregate functions. Thus, they are defined
over a domain that
contains only a single value, the empty tuple. Hence, $\COUNT$
and $\PRTY$ are also singleton-determining aggregation functions.
However
$\CNTD$ is not singleton-determining aggregation functions.

\begin{theorem}[Equivalence of Quasilinear
  Queries]\label{T:singleton-determining:proper} 
Let $\alpha$ be an aggregation function. 
Then, the following conditions are equivalent:
\begin{enumerate}
\item $\alpha$ is singleton-determining;
    \label{cond:1}
\item $\Lin\alpha$ is proper; 
    \label{cond:new}
\item $\QL\alpha$ is proper.
    \label{cond:2}
\end{enumerate}
\end{theorem}

\begin{proof}
The direction \quotes{\eqref{cond:new} $\Rightarrow$ \eqref{cond:2}} 
holds by Theorem~\ref{thm-negation:positive:isomorphism}. 
Clearly,
\quotes{\eqref{cond:2} $\Rightarrow$ \eqref{cond:new}} holds 
since $\Lin\alpha \subseteq \QL \alpha$. 
Thus, we need only show that 
\quotes{\eqref{cond:1} $\Rightarrow$ \eqref{cond:new}} and
\quotes{\eqref{cond:new} $\Rightarrow$ \eqref{cond:1}}. 

\medskip
\quotes{\eqref{cond:1} $\Rightarrow$ \eqref{cond:new}}\ \ 
Suppose that $\alpha$ is a singleton-determining aggregation function. 
We show that $\Lin\alpha$ is proper.
To this end, 
let $q(\tpl s,\alpha(\tpl t))\qif A$ and $q'(\tpl s,\alpha(\tpl
t))\qif A'$ be
satisfiable reduced  linear $\alpha$-queries. 
Suppose that $q\equiv q'$. We will show that $q$ and $q'$ are isomorphic. 

In~\cite{Chandra:Merlin-Conjunctive:Queries-STOC} it has been shown that 
positive linear non-aggregate queries without comparisons are set-equivalent 
if and only if they are isomorphic. This still holds even if the
queries have comparisons.\footnote{We are not aware that this result
has been published, but it appears in the extended version
  of~\cite{Nutt:Et:Al-Equivalences:Among:Aggregate:Queries-PODS}.} 
We associate with $q$ a non-aggregate query $\naproj q$, 
called the 
 {\em non-aggregate projection\/} of $q$, which is derived from $q$ by
 simply removing the aggregate term from the head of $q$. Thus,
 $\naproj q$ has the form
 \begin{eqnarray*}
 \label{non-aggregate:counterpart}
         \naproj q(\tpl s) \qif A.
 \end{eqnarray*}

Since $q\equiv q'$, they return values for the same grouping tuples. Thus,
$\naproj q$ is set-equivalent to $\naproj q'$. Hence, $\naproj q$ is
isomorphic to 
$\naproj q'$. Let $\theta$ be the isomorphism from $\naproj q'$ to
$\naproj q$. If $\alpha$ is a nullary aggregation function, then
$\theta$ is an isomorphism from $q'$ to $q$. Suppose that $\alpha$ is
not a nullary aggregation function.

Let $\gamma$ be an instantiation of the terms in $q$
that satisfies the comparisons in $q$ and maps each term to a
different value. 
We construct a database $\D$ out  of $q$ by applying
$\gamma$ to the relational part of $q$. 

Clearly, the only satisfying
assignment of $q$ to the constants in $\D$ is exactly $\gamma$. Thus,
$q$ retrieves $(\gamma (\tpl s), \alpha(\gamma (\tpl t)))$. The only
satisfying assignment of $q'$ is $\gamma \compose \theta$. Therefore,
$q'$ returns $(\gamma \compose \theta (\tpl s), \alpha(\gamma \compose
\theta (\tpl t)))$. Note that since $\theta$ is an isomorphism from
$q'$ to $q$, it holds that $\gamma \compose \theta(\tpl s)= \gamma
(\tpl s)$. 

Recall that $\alpha$ is a singleton-determining aggregation
function. Therefore, $\alpha(\gamma \compose \theta (\tpl t))) =
\alpha(\gamma (\tpl t)))$ if and only if $\gamma \compose \theta (\tpl t)
= \gamma (\tpl t)$. The instantiation $\gamma$ is an injection, thus
$\gamma \compose \theta (\tpl t)
= \gamma (\tpl t)$ if and only if $\theta (\tpl t) = \tpl t$. This must
hold since $q\equiv q'$. Therefore, $\theta$ is an isomorphism from
$q$ to $q'$.

\medskip
\quotes{\eqref{cond:new} $\Rightarrow$ \eqref{cond:1}}\ \ 
Suppose that $\alpha$ is not 
a singleton-determining aggregation function. 
We show that $\Lin\alpha$ is not proper.
To this end, 
we create linear $\alpha$-queries $q$ and $q'$ such that 
$q\equiv q'$, but $q$ and $q'$ are not isomorphic. 

Since $\alpha$ is not a singleton-determining aggregation function,
there are singleton bags $B = \bag{d}$, and $B' = \bag{d'}$ such that
$d\neq d'$ and $\alpha(B) = \alpha(B')$. We define the queries
\begin{align*}
   q(\alpha(d))  & \qif p(d) \AND p(d') \\
  q'(\alpha(d')) & \qif p(d) \AND p(d').
\end{align*}
Clearly $q$ and $q'$ are not isomorphic, but they are equivalent.
\end{proof}

\begin{corollary}[Equivalence and Isomorphism]
The classes of quasilinear 
$\MAX$, $\TOPTWO$, $\COUNT$, $\SUM$, $\PROD$, $\PRTY$ and  $\AVG$
queries are proper.
\end{corollary}

\begin{proof}
This result follows from the fact that all the aggregation functions
above are singleton-determining and from
Theorem~\ref{T:singleton-determining:proper}. 
\end{proof}

For $\CNTD$ a similar result can be shown for common cases. 

\begin{theorem}[Equivalence of Quasilinear Count-Distinct Queries]
Let $q$ and $q'$ be satisfiable reduced quasilinear $\CNTD$-queries.
Moreover, suppose that
\begin{itemize}
  \item the comparisons in $q$ and $q'$ use only $\leq$, $\geq$ {\em and}
  \item $q$ and $q'$ either range over the rational numbers 
        {\em or\/} do not have constants.
\end{itemize}
Then $q$ and $q'$ are equivalent if and only if they are isomorphic.
\end{theorem}

\begin{proof}
This follows directly from the fact that such queries, when positive,
are equivalent if and only if they are
isomorphic~\cite{Nutt:Et:Al-Equivalences:Among:Aggregate:Queries-PODS}
and from Theorem~\ref{thm-negation:positive:isomorphism}. 
\end{proof}

Since isomorphism of quasilinear queries can be checked in polynomial
time, we derive the following complexity result.

\begin{corollary}[Polynomiality]
The equivalence problem for the class of quasilinear $\alpha$-queries
is decidable in polynomial time if $\alpha$ is one of the aggregation
functions $\MAX$, $\TOPTWO$, $\COUNT$, 
$\SUM$, $\PROD$, $\PRTY$, or $\AVG$ and for common $\CNTD$-queries.
\end{corollary}

\eat{
{\bigskip

\noindent
\sl COMMENT: I haven't done anything on this section.

with this section, I am not completely sure whether
\begin{itemize}
\item the theorem is correct;
\item whether it can be generalized (perhaps after a suitable 
      modification if it not yet correct).
\end{itemize}

A generalization could consist in formulating it for all monoidal
aggregation functions and/or for disjunctive queries.
At least we should make sure that it holds also for $\PRTY$ and
$\TOPTWO$.
For the disjunctive queries it may be that for group queries we still
have isomorphism.  

For the idempotent queries we surely don't have simple isomorphism,
but we may find for every maximally general condition in one query an
isomorphic condition in the other one.  By \quotes{a maximally
general} condition I mean a condition that is not properly contained
in another condition of the query.
}

}


\section{Conclusion}
Necessary and complete conditions for the decidability of bounded
equivalence of disjunctive aggregate queries with negation have been
presented.
This problem has been shown to be decidable for a wide class of
aggregation functions.
Equivalence of aggregate queries with negation has been reduced to a
special case of bounded equivalence, called local equivalence, for
decomposable aggregation functions.
We have also shown that equivalence can be decided in polynomial time
for the common case of quasilinear
queries. 

Novel proof techniques have been presented. 
One example is the application of the Principle of Inclusion and Exclusion 
to the case of group aggregation functions. 
Our results are couched in terms of abstract characterizations
of aggregation functions. 
Thus, the results presented are easily extendible to 
additional aggregation functions. In
Table~\ref{tab:properties:aggregation:functions} we 
summarize the properties that hold for  each of the aggregation functions
considered in this paper. 
Table~\ref{tab:properties:classes:queries} shows our decidability
results for these aggregation functions.

\begin{table}[t]
\begin{center}
\fbox{
\begin{tabular}{rcccc}
\rule[-8pt]{0pt}{22pt}%
\parbox{50pt}{} & { Shiftable} &
{Order-Decidable} & {Decomposable} & {Singleton-Determining} \\
\hline\rule{0pt}{14pt}%
$\COUNT$ & $\surd$ &  $\surd$ &  $\surd$ &  $\surd$ \\\rule{0pt}{14pt}%
$\MAX$   & $\surd$ &  $\surd$ &  $\surd$ &  $\surd$ \\\rule{0pt}{14pt}%
$\SUM$   &         &  $\surd$ &  $\surd$ &  $\surd$ \\\rule{0pt}{14pt}%
$\PROD$  &         &  $\surd$ &  over $\ratnozero$
&  $\surd$ \\ \rule{0pt}{14pt}%
$\TOPTWO$& $\surd$ &  $\surd$ &  $\surd$ &  $\surd$ \\\rule{0pt}{14pt}%
$\AVG$   &         &  $\surd$ &          &  $\surd$ \\\rule{0pt}{14pt}%
$\CNTD$  & $\surd$ &  $\surd$ &          &          \\\rule{0pt}{14pt}%
$\PRTY$  & $\surd$ &  $\surd$ &  $\surd$ &  $\surd$ 
\end{tabular}}
\end{center}
\caption{Properties of aggregation
  functions\label{tab:properties:aggregation:functions}} 
\end{table}

\begin{table}[t]
\begin{center}
\fbox{
\begin{tabular}{rccc}
\parbox{50pt}{} & { Decidability of} &
{ Decidability of} & {Equivalence is Isomorphism} \\
\rule[-8pt]{0pt}{12pt}%
\parbox{50pt}{} & { Bounded Equivalence} &
{Equivalence } & {for Quasilinear Queries} \\
\hline\rule{0pt}{14pt}%
$\COUNT$ &   $\surd$ &  $\surd$ &  $\surd$ \\\rule{0pt}{14pt}%
$\MAX$   &   $\surd$ &  $\surd$ &  $\surd$ \\\rule{0pt}{14pt}%
$\SUM$   &   $\surd$ &  $\surd$ &  $\surd$ \\\rule{0pt}{14pt}%
$\PROD$  &   $\surd$ &  $\surd$ &  $\surd$ \\ \rule{0pt}{14pt}%
$\TOPTWO$&   $\surd$ &  $\surd$ &  $\surd$ \\\rule{0pt}{14pt}%
$\AVG$   &   $\surd$ &          &  $\surd$ \\\rule{0pt}{14pt}%
$\CNTD$  &   $\surd$ &          & special cases \\\rule{0pt}{14pt}%
$\PRTY$  &   $\surd$ &  $\surd$ &  $\surd$ 
\end{tabular}
}
\end{center}
\caption{Properties of classes of queries
\label{tab:properties:classes:queries}} 
\end{table}

Bag-set semantics has been introduced
in~\cite{Chaudhuri:Vardi-Real:Conjunctive:Queries-PODS} 
to give a formal account of the way in which SQL queries are executed,
which do not return a set of tuples but a multiset.
It is easy to see that two non-aggregate queries are equivalent under
bag-set semantics if and only if the aggregate queries obtained by
adding the function $\COUNT$ are equivalent.
Thus, our results on $\COUNT$-queries directly carry over to
non-aggregate queries that are evaluated under bag-set semantics.
This is a significant contribution to the understanding of SQL
queries.
Moreover, these results can easily be extended to 
non-aggregate queries evaluated under {\em bag semantics}
\cite{Chaudhuri:Vardi-Real:Conjunctive:Queries-PODS, 
      Ioannidis:Ramakrishnan-Beyond:Relations:As:Sets-TODS},
thereby, solving an additional open problem.

Concepts seemingly similar to the ones introduced in the present paper
have been investigated in \cite{Hella:Et:Al-Logics-Aggregate-Operators:LICS}.
In particular, the authors considered aggregation functions defined in
terms of commutative monoids.
However, the purpose of that research was to study the expressivity of
logics that extend first-order logic by aggregation. 
In \cite{Hella:Et:Al-Logics-Aggregate-Operators:LICS} it is shown that
formulas in those extended logics are Hanf-local and Gaifman-local. 
Intuitively, this means that whether or not 
a formula is true for a tuple $\tpl d$ in a structure, 
depends only on that part of the structure that is ``close'' to $\tpl d$.  
A class of formulas that is Hanf- or Gaifman-local need not be
decidable.
In addition, the authors only considered monoids over the rational numbers,
which excludes functions such as $\TOPK$ and $\PRTY$.

We leave for future research the problem of deciding equivalence among 
$\AVG$ and $\CNTD$ queries as well as equivalence of 
aggregate queries with a {\tt HAVING} clause. Finding tight upper and
lower bounds for equivalence, as well as 
the adaptation of our results to the view usability problem 
are other important open problems. 


\bibliographystyle{abbrv}
\bibliography{strings,%
             literature}

\begin{thebibliography}{10}

\bibitem{Aho:Et:Al-Efficient:Optimization-TODS}
A.~Aho, Y.~Sagiv, and J.~Ullman.
\newblock Efficient optimization of a class of relational expressions.
\newblock {\em ACM Transactions on Database Systems}, 4(4):435--454, 1979.

\bibitem{Chandra:Merlin-Conjunctive:Queries-STOC}
A.~Chandra and P.~Merlin.
\newblock Optimal implementation of conjunctive queries in relational
  databases.
\newblock In {\em Proc. 9th Annual ACM Symposium on Theory of Computing}, pages
  77--90. ACM Press, May 1977.

\bibitem{Chaudhuri:Vardi-Real:Conjunctive:Queries-PODS}
S.~Chaudhuri and M.~Vardi.
\newblock Optimization of real conjunctive queries.
\newblock In {\em Proc. 12th Symposium on Principles of Database Systems},
  Washington (D.C., USA), May 1993. ACM Press.

\bibitem{Negation}
S.~Cohen, W.~Nutt, and S.~Sagiv.
\newblock Equivalences among aggregate queries with negation.
\newblock In {\em Proc. 20th Symposium on Principles of Database Systems},
  pages 215--226, Santa Barbara (California, USA), May 2001. ACM Press.

\bibitem{Cohen:Et:Al-Rewriting:Aggregate:Queries-PODS}
S.~Cohen, W.~Nutt, and A.~Serebrenik.
\newblock Rewriting aggregate queries using views.
\newblock In C.~Papadimitriou, editor, {\em Proc. 18th Symposium on Principles
  of Database Systems}, Philadelphia (Pennsylvania, USA), May 1999. ACM Press.

\bibitem{Grumbach:Rafanelli:Tininini-Querying:Aggregate:Data-PODS}
S.~Grumbach, M.~Rafanelli, and L.~Tininini.
\newblock Querying aggregate data.
\newblock In C.~Papadimitriou, editor, {\em Proc. 18th Symposium on Principles
  of Database Systems}, pages 174--183, Philadelphia (Pennsylvania, USA), May
  1999. ACM Press.

\bibitem{Hella:Et:Al-Logics-Aggregate-Operators:LICS}
L.~Hella, L.~Libkin, J.~Nurmonen, and L.~Wong.
\newblock Logics with aggregate operators.
\newblock In {\em Proc. 14th IEEE Symposium on Logic in Computer Science},
  pages 35--44, Trento (Italy), July 1999. IEEE Computer Society Press.

\bibitem{Ioannidis:Ramakrishnan-Beyond:Relations:As:Sets-TODS}
Y.~Ioannidis and R.~Ramakrishnan.
\newblock Beyond relations as sets.
\newblock {\em ACM Transactions on Database Systems}, 20(3):288--324, 1995.

\bibitem{Johnson:Klug-Optimizing:Conjunctive:Queries-SIAMJ}
D.~Johnson and A.~Klug.
\newblock Optimizing conjunctive queries that contain untyped variables.
\newblock {\em SIAM Journal on Computing}, 12(4):616--640, 1983.

\bibitem{Kreisel-Krivine:Model-Theory}
G.~Kreisel and J.~L. Krivine.
\newblock {\em Elements of Mathematical Logic: Model Theory}.
\newblock North Holland (Amsterdam), 1967.

\bibitem{Levy:Sagiv-Queries:Independent:Updates-VLDB}
A.~Levy and Y.~Sagiv.
\newblock Queries independent of updates.
\newblock In {\em Proc. 19th International Conference on Very Large Data
  Bases}, pages 171--181, Dublin (Ireland), Aug. 1993. Morgan Kaufmann
  Publishers.

\bibitem{Levy:Sagiv-Sem:Query:Opt-PODS}
A.~Levy and Y.~Sagiv.
\newblock Semantic query optimization in datalog programs.
\newblock In {\em Proc. 14th Symposium on Principles of Database Systems},
  pages 163--173, San Jose (California, USA), Proc. 14th Symposium on
  Principles of Database Systems 1995. ACM Press.

\bibitem{Nutt:Et:Al-Equivalences:Among:Aggregate:Queries-PODS}
W.~Nutt, Y.~Sagiv, and S.~Shurin.
\newblock Deciding equivalences among aggregate queries.
\newblock In J.~Paredaens, editor, {\em Proc. 17th Symposium on Principles of
  Database Systems}, pages 214--223, Seattle (Washington, USA), June 1998. ACM
  Press.
\newblock Long version as Report of Esprit LTR DWQ.

\bibitem{Presburger}
M.~Presburger.
\newblock {\"U}ber die {V}ollst{\"{a}}ndigkeit eines gewissen {S}ystems der
  {A}rithmetik ganzer {Z}ahlen, in welchem die {A}ddition als einzige
  {O}peration hervortritt.
\newblock In {\em 1. Kongres matematyk\'{o}w krajow slowia\'{n}skich}, pages
  92--101, Warsaw, 1929.

\bibitem{Sagiv:Yannakakis-Union:And:Difference-JACM}
Y.~Sagiv and M.~Yannakakis.
\newblock Equivalence among relational expressions with the union and
  difference operators.
\newblock {\em J. ACM}, 27(4):633--655, 1981.

\bibitem{Ullman-Principles:Of:DB:And:KB:Systems}
J.~D. Ullman.
\newblock {\em Principles of Database and Knowledge-Base Systems}, volume~I.
\newblock Computer Science Press, 1988.

\bibitem{Van:Der:Meyden-Linearly:Ordered:Domains-PODS}
R.~van~der Meyden.
\newblock The complexity of querying indefinite data about linearly ordered
  domains.
\newblock In {\em Proc. 11th Symposium on Principles of Database Systems},
  pages 331--345, San Diego (California, USA), May 1992. ACM Press.

\end{thebibliography}

\end{document}